\documentclass[aps,floatfix,showpacs,superscriptaddress,nofootinbib,natbib]{revtex4}
\usepackage{epsfig}
\usepackage[english]{babel}
\usepackage{pgf}
\usepackage{amsfonts,amsmath,amssymb,enumerate,array,amssymb}
\usepackage{enumitem}
\usepackage{mathtools} 
\usepackage{multirow} 
\usepackage[latin1]{inputenc}
\usepackage{bm}
\usepackage{fixmath}
\usepackage{amsbsy}
\usepackage{graphicx}
\usepackage{capt-of}
\usepackage{dcolumn}
\usepackage{upgreek}
\usepackage{float}
\usepackage{subfig}
\usepackage[format=plain,justification=centerlast]{caption}
\usepackage{bbm}        

\usepackage{layouts}
\usepackage{natbib}
\usepackage{abrev-jour-long}

\usepackage{pdfpages}
\usepackage{color}
\usepackage{graphicx,epstopdf}
\usepackage[colorlinks=true,
linkcolor=red,
urlcolor=purple,
citecolor=blue,hyperindex]{hyperref}

\begin{document}

	\title{Gravity Localization on Intersecting Thick Braneworlds}

	\author{Henrique Matheus Gauy}\email[]{henmgauy@df.ufscar.br}

	\author{Alex E. Bernardini}\email[]{alexeb@ufscar.br}

	\affiliation{Departamento de F\'isica, Universidade
		Federal de S\~{a}o Carlos, S\~ao Carlos, 13565-905 SP, Brazil}

	
	\begin{abstract}
		The localization of gravity for $(5+1)$-dimensional intersecting thick braneworld models is thoroughly investigated. Departing from preliminary results for five independent proposals, from $I$ to $V$, gravity is shown to be localized in the brane. In particular, for models from $I$ to $III$, only the zero modes are analytically determined. On the other hand, for models $IV$ and $V$, massive modes are all obtained: a finite number of massive states is identified for model $IV$, while an infinite but discrete number of massive states bounded from below is exhibited by model $V$. Considering that the discreteness of the graviton modes implies that gravity does not propagate in the co-dimensions, the naked singularities of models $IV$ and $V$, at edges of space, are made harmless. To conclude, the phenomenological implications for sphere models, constructed out of models $III$ and $IV$, are also discussed. According to such an overall classification, if the internal space is a sphere, model $IV$ is shown to exhibit normalizable modes. More relevantly, for the sphere model assembled out of model $III$, the subtle property of reproducing a consistent Newtonian limit is identified.
	\end{abstract}
	
	\date{\today}
	
	\maketitle

\section{Introduction}

The idea that our universe is embedded in some higher dimensional space has been interpreted as a viable paradigm since it was seminally introduced by Randall and Sundrum (RS) \cite{RS-1,RS-2} through a self-gravitating brane model which would work as an alternative proposal to the so-called compactified models (like the ADD models \cite{ArkaniHamed1998a}). In such scenarios, the curvature of space engenders a localization mechanism which admits a suppression of the higher dimensional terms, thus recovering, within certain limits, the Newton's gravitation theory in the brane. Given that RS models rely on thin branes, and the standard model fields are confined in a thin slice of space, a more realistic framework suggests that the matter fields ought be smeared over extra dimensions. Mostly driven by a system of dark scalar fields, thick braneworlds hence replace the thin brane with some topological defect, therefore admitting some lump-like (non-topological) defect solution for the warp factor \cite{Bernardini2016,Almeida2009,Bazeia2004,Dzhunushaliev2010,Bazeia2002,DeWolfe2000,Ahmed2013,Gremm2000a,Gremm2000,Kehagias2001,Kobayashi2002,Bronnikov2003,BAZEIA2009b,BarbosaCendejas2013,ZHANG2008,Melfo2003,Bazeia2009a,Koley2005,Bazeia2014,Chinaglia2016,Bernardini2013}. In this case, the non-topological nature of curvature begets the same localization mechanism of the RS models for gravitational and matter fields \cite{Bernardini2016,Almeida2009,Bazeia2004,Dzhunushaliev2010,Bazeia2002,DeWolfe2000,Ahmed2013,Gremm2000a,Gremm2000,Kehagias2001,Kobayashi2002,Bronnikov2003,BAZEIA2009b,BarbosaCendejas2013,ZHANG2008,Melfo2003,Bazeia2009a,Koley2005,Bazeia2014,Chinaglia2016,Bernardini2013,HERRERAAGUILAR2010,Guo2013}, which leads to straightforward astrophysical and cosmological implications \cite{Hall1999,Ahmed2014,Csaki2000a,Rubakov1983a,Binetruy2000,Binetruy2000a,Cline1999,Csaki1999,Csaki2000b,Flanagan2000,Kanti2000,Kanti1999,Bazeia2008,George2009,Kadosh2012,Kim2000,Binetruy2000b,Bowcock2000,Guha2018,Mukohyama2000,Chung1999,Ida2000,Chung2000,Mukohyama2000a,MERSINI2001,Bernardini2014,Casadio2014}.

Besides the above mentioned five dimensional scenarios, six dimensional models have also been considered \cite{Csaki2000,ArkaniHamed2000,PARAMESWARAN200754,Liu2007,Dzhunushaliev2008,Koley2006,Singleton2004,Gherghetta2000,Park2003,Dzhunushaliev2009,Multamaeki2002,Gregory2000,Cohen1999} and stimulated the emergence of novel classes of six dimensional intersecting braneworlds \cite{Gauy2022}. In the context of the background results that shall be recovered along this work, 
six dimensional localized branes generated by two intersecting scalar fields (where the warp factor and the components of the metric are assumed as separable functions) have recently been categorized into five different analytical models \cite{Gauy2022}, classified from $I$ to $V$. 
In this case, a twofold-warp factor constructed from two intersecting warp factors results into a bulk configuration where the brane localization is driven by two crossing scalar fields.

Considering that any braneworld model should recover standard four dimensional physics, at least within reasonable limits \cite{Antoniadis1998}, and assuming that all physical fields must be localized in a brane-like region of space, 
it is natural to suppose that such a two co-dimensional thick brane-world admits the localization of gravity for all intersecting thick braneworld configurations, from $I$ to $V$, in correspondence with the results from Ref.~\cite{Gauy2022}. 
Besides the inherent classical aspects, from systematic metric perturbations over the braneworld \cite{RS-1,RS-2}, which renders a linearized formulation that encompass non-linear aspects of gravity, the equations for the metric fluctuations can be reduced to time independent Schr\"odinger-like equations \cite{RS-1,RS-2,Csaki2000,Gremm2000a,Gremm2000}, which are then subjected to some separation of variables technique that results into the quantum mechanical analogue problem.

The quantum mechanical aspects will then be further extended to the interpretation of the associated wave functions, solutions of such Schr\"odinger-like equation, which are not only relevant for the localization on the brane, but also for identifying the Newtonian limit.
Recovering the Newtonian gravity in the brane, up to a sensitive limit, is indeed an important test for braneworld models. One may determine the Newtonian limit by assuming that perturbations of the metric are generated by a particle of mass $\mathcal{M}$. As a first approach, the effective coupling between the massive particle with gravity becomes dependent on the normalization of the graviton modes and on its value at the position of the particle (which is generically placed at the maximum of the wave functions of the zero mode). However, a more realistic description would consider any particle to be smeared over the co-dimensions \cite{Csaki2000,Brandhuber1999}, in contrast to a Dirac delta localization. In this case, the gravitational constant would no longer depend on the value of the graviton modes at the position of the particle, but would be determined by an average value of the matter distribution with the graviton modes. 
For the proposal considered here, the simplified stratagem of a Dirac delta distribution is enough, since one can at least determine the correct gravitational constant, even if failing to ascertain the correct gravitational interaction\footnote{Generically, the Newtonian limit of some braneworld should append some modifications to the gravitational interaction. To avoid contradictions with experimental data one expects the corrections to be insignificant above some scale, thus recovering Newtonian gravity.}.

Broadly speaking, the pertinence of considering six dimensional braneworlds resides in obtaining Schr\"odinger-like linearized equations for a two (and not only one) dimensional curved background. On one hand, intersecting thick branes allow for a further separation of the co-dimensional variables for a subset of our models (from $III$ to $V$). On the other hand, for another subset (models $I$ and $II$), the variables are not separable unless the graviton mass is null. Furthermore, for two of the models which shall be considered here, $IV$ and $V$, Schr\"odinger-like equations solutions are analytically defined for the entire spectrum of gravitons. In particular, this provides a wide range for the mass spectrum. For models $I$ to $III$, however, only zero modes can be analytically extracted. Regardless, some remarkable features are still accomplished from the Newtonian limit for all these models, from $I$ to $V$, as they shall be scrutinized along this paper.

The paper is thus organized as follows. In Sec.~\ref{preliminaries}, a brief review of the solutions for $(5+1)$-dimensional intersecting thick braneworlds driven by two scalar fields, in correspondence to models from $I$ to $V$ (cf. Ref.~\cite{Gauy2022}) is presented. Sec.~\ref{fluctuations} is devoted to determine the equations for the gravitational perturbations for a general background, which straightforwardly maps the quantum mechanical analogue problem that describes the localization and spectrum of the graviton modes. Finally, in Sec.~\ref{quantum}, the Schr\"odinger equation for each intersecting thick braneworld model is yielded and their corresponding spectrum of gravitons as well as their Newtonian limit are obtained. Our conclusions are drawn in Sec.~\ref{Concl} pointing out the most relevant features and delineating comparisons between the proposed models.

\section{(5+1)-Dimensional Intersecting Brane-Worlds Preliminaries}\label{preliminaries}

Let the braneworld be a six dimensional manifold $\mathbb{E}^{6}$ that is, as a set, equivalent to the product space $\mathbb{M}^{4}\times \mathbb{B}^{2}$, where $\mathbb{M}^{4}$ and $\mathbb{B}^{2}$ are manifolds of four and two dimensions, respectively. The following ansatz is assumed for the metric of $\mathbb{E}^{6}$ \cite{Gauy2022}:
	\begin{equation}
	\boldsymbol{g}=e^{-2A}\omega_{\mu\nu}\mathrm{d}x^{\nu}\mathrm{d}x^{\mu}+ e^{-2f}\mathrm{d}u^{2}+e^{-2h}\mathrm{d}v^{2},\label{bundlemetric}
	\end{equation}
	where $A$ is the warp factor, $\boldsymbol{\omega}$ is the metric of the space-time $\mathbb{M}^{4}$, $\boldsymbol{\omega}:\mathbb{M}^{4}\rightarrow\mathcal{T}^{\left(0,2\right)}\mathbb{M}^{4}$, and $e^{-2f}$ and $e^{-2h}$ are the components of metric $\boldsymbol{\sigma}$ of the internal space, $\mathbb{B}^{2}$, $\boldsymbol{\sigma}:\mathbb{B}^{2}\rightarrow\mathcal{T}^{\left(0,2\right)}\mathbb{B}^{2}$\footnote{To clear up the notation, Greek indices ($\mu$, $\nu$,...) are valued in the set $\{0,1,2,3\}$, uppercase Latin indices ($M$, $N$,...) in the set $\{0,1,2,3,4,5\}$, lowercase Latin indices ($m$, $n$, $i$, $j$,...), representing the bulk co-dimension, in the set $\{4,5\}$, and the labels $x^{4}=u$ and $x^{5}=v$, representing the choice of coordinates for the co-dimensions $(\mathbb{B}^{2})$. The use of notation $T_{45}\equiv T_{uv}$ whenever suited, indicates that $``4 \leftrightarrow u''$ and $``5 \leftrightarrow v''$. Derivatives, whenever suited, will be represented by a {\em comma}, i.e. $f_{,\mu}\leftrightarrow \partial f/\partial x^{\mu}$. Finally, tensors when being referred to its (abstract) entirety will be in boldface, as $\boldsymbol{g}$, but its components will be cast in regular font, as $g_{\mu\nu}$.}. Here $A:\mathbb{B}^{2}\rightarrow \mathbb{R}$, which means that $A=A(u,v)$; $\omega_{\mu\nu}:\mathbb{M}^{4}\rightarrow \mathbb{R}$; $f:\mathbb{B}^{2}\rightarrow \mathbb{R}$; and $h:\mathbb{B}^{2}\rightarrow \mathbb{R}$.
	
	The action of gravity is the usual Einstein-Hilbert action in six dimensions minimally coupled to two canonical real scalar fields,
	\begin{equation}
	S=\int\mathrm{d}^{6}x\sqrt{-\mathrm{g}} \left[2{M}^{4}\,R-\left(\frac{g^{MN}}{2}\phi_{,M}\phi_{,N}+\frac{g^{MN}}{2}\zeta_{,M}\zeta_{,N}+\mathcal{V}\right)\right],
	\end{equation}
	where $R$ is a six dimensional Ricci scalar, $\phi:\mathbb{B}^{2}\rightarrow \mathbb{R}$ ($\phi{\equiv}\phi(u,v)$), $\zeta:\mathbb{B}^{2}\rightarrow \mathbb{R}$ ($\zeta{\equiv}\zeta(u,v)$), $\mathcal{V}$ is some function of $\phi$ and $\zeta$, and $\mathrm{g}=\det{\left(g_{MN}\right)}$.

	For intersecting braneworlds, one assumes that the scalar fields are functions of different variables, i.e. $\phi_{,v}=0$ and $\zeta_{,u}=0$, and the metric components are separable functions, i.e. $A=\hat{A}(v)+\tilde{A}(u)$, $f=\hat{f}(v)+\tilde{f}(u)$ and $h=\hat{h}(v)+\tilde{h}(u)$. In this case, it follows from Einstein equations that the metric must take one of the two forms \cite{Gauy2022}
	\begin{equation}
	\boldsymbol{g}=e^{-2\tilde{A}}\omega_{\mu\nu}\mathrm{d}x^{\mu}\mathrm{d}x^{\nu}+e^{-2\tilde{f}}\mathrm{d}u^{2}+e^{-2\tilde{h}}\mathrm{d}v^{2},\label{stringlikedefect}
	\end{equation}
	\begin{equation}
	\boldsymbol{g}=e^{-2\hat{A}}e^{-2\tilde{A}}\omega_{\mu\nu}\mathrm{d}x^{\mu}\mathrm{d}x^{\nu}+e^{-2p\hat{A}}e^{-2\tilde{f}}\mathrm{d}u^{2}+e^{-2\hat{h}}e^{-2\left(1-p\right)\tilde{A}}\mathrm{d}v^{2}.\label{intersectingmetric}
	\end{equation}
	Metric Eq.~\eqref{stringlikedefect} corresponds to a string like defect \cite{Koley2006,Gherghetta2000,Park2003,PARAMESWARAN200754,Singleton2004,Multamaeki2002} and metric Eq.~\eqref{intersectingmetric}, which shall be considered from this point, can be re-organized in terms of the so-called models from $I$ to $V$ (cf. \cite{Gauy2022}), as they are described by
	\begin{align}
	\boldsymbol{g}^{I}&=\sqrt{\cosh \left(2c_{u}u\right)\left|\cos \left(2c_{v}v\right)\right|}\eta_{\mu\nu}\mathrm{d}x^{\mu}\mathrm{d}x^{\nu}+\sqrt{\frac{\left|\cos \left(2c_{v}v\right)\right|^{p} }{\cosh^{p-1} \left(2c_{u}u\right)}}\left(\mathrm{d}u^{2}+\mathrm{d}v^{2}\right)\text{, }p>3;\label{metricI}
	\\
	\boldsymbol{g}^{II}&=\sqrt{\left|\cos \left(2c_{u} u\right)\cos \left(2c_{v}v\right)\right|}\eta_{\mu\nu}\mathrm{d}x^{\mu}\mathrm{d}x^{\nu}+\sqrt{\frac{\left|\cos \left(2c_{v} v\right)\right|^{p}}{\left|\cos \left(2c_{u} u\right)\right|^{p-1}}}\left(\mathrm{d}u^{2}+\mathrm{d}v^{2}\right)\text{, }\frac{1}{2}\leq p\leq 3;\label{metricII}
	\\
	\boldsymbol{g}^{III}&= \sqrt{\left|\cos\left(2\sqrt{C}v\right)\right|}e^{-2\tilde{A}}\eta_{\mu\nu}\mathrm{d}x^{\mu}\mathrm{d}x^{\nu}+e^{-2\tilde{f}}\mathrm{d}u^{2}+e^{-2\tilde{A}}\mathrm{d}v^{2};\label{metricIII}
	\\
	\boldsymbol{g}^{IV}&= \frac{4\Lambda}{3C}\cos^{2}\left(\frac{\sqrt{C} v}{2}\right) e^{-2\tilde{A}}\omega^{+}_{\mu\nu}\mathrm{d}x^{\mu}\mathrm{d}x^{\nu}+e^{-2\tilde{f}}\mathrm{d}u^{2}+e^{-2\tilde{A}}\mathrm{d}v^{2};\label{metricIV}
	\\
	\boldsymbol{g}^{V}&=\cos^{2/3}\left(\sqrt{3\left|\Lambda\right|}v\right)e^{-2\tilde{A}}\omega^{-}_{\mu\nu}\mathrm{d}x^{\mu}\mathrm{d}x^{\nu}+e^{-2\tilde{f}}\mathrm{d}u^{2}+\cos^{2/3}\left(\sqrt{3\left|\Lambda\right|}v\right)e^{-2\tilde{A}}\mathrm{d}v^{2},\label{metricV}
	\end{align}
	where $c_{u}$, $c_{v}$, $C$ and $\Lambda$ are separation constants; $\omega^{\pm}_{\mu\nu}$ represents solutions of the equation $\mathcal{R}_{\mu\nu}=\Lambda \omega_{\mu\nu}$, with either positive (for $\omega^{+}_{\mu\nu}$) or negative (for $\omega^{-}_{\mu\nu}$) values of $\Lambda$; and $\mathcal{R}_{\mu\nu}$ represents the Ricci tensor compatible with $\omega_{\mu\nu}$. The warp factor $\tilde{A}$ of models $III$, $IV$ and $V$ are not strictly specified by Einstein equations and can be left free of constraints. The localization of gravitons and the Newtonian limit can thus be settled for the above set of geometries, from $I$ to $V$.

\section{Gravitational Fluctuations}\label{fluctuations}

A basic requirement of any braneworld model is that it reproduces, up to a sensible limit, Newtonian gravity \cite{RS-1,RS-2,Csaki2000,Gremm2000a,Gremm2000}. The Newtonian limit of any braneworld model can be studied within the framework of linearized gravity, where it is assumed that the spacetime can be described by small fluctuations about a given background. Following similar arguments as outlined by \cite{Csaki2000}\footnote{Extending this line of reasoning in order to account for a curved four dimensional space \cite{Isaacson1968}.}, tensorial perturbations of \eqref{bundlemetric} are written as
\begin{equation}
\bar{\boldsymbol{g}}=e^{-2A}\left[{\omega}_{\mu\nu}+\varpi_{\mu\nu}\right]\mathrm{d}x^{\mu}\mathrm{d}x^{\nu}+\sigma_{ij}\mathrm{d}y^{i}\mathrm{d}y^{j},\label{metricperturbation}
\end{equation}
where $\varpi_{\mu\nu}$ are functions such that $\varpi_{\mu\nu}:\mathbb{M}^{4}\times \mathbb{B}^{2}\rightarrow \mathbb{R}$ (i.e. $\varpi_{\mu\nu}(x^{\rho},u,v)$). The Einstein tensor ($\bar{G}_{MN}$) compatible with $\bar{\boldsymbol{g}}$ is then expressed, up to the first order in $\varpi_{\mu\nu}$, by
\begin{align}
\bar{G}_{MN}&={G}_{MN}+\delta G_{MN}\nonumber
\\
&={G}_{MN}+\frac{1}{2}\hat{g}^{PK}\hat{\nabla}_{P}\hat{\nabla}_{N}{\varpi}_{MK}+\frac{1}{2}\hat{g}^{PK}\hat{\nabla}_{P}\hat{\nabla}_{M}{\varpi}_{KN}-\frac{1}{2}\hat{\square}{\varpi}_{MN}-\frac{1}{2}\hat{g}^{PK}\hat{\nabla}_{M}\hat{\nabla}_{N}{\varpi}_{PK}\nonumber
\\
&\qquad\qquad-\frac{1}{2}\varpi_{MN}\hat{R}+\frac{1}{2}\hat{g}_{MN}\varpi_{PK}\hat{R}^{PK}-\frac{1}{2}\hat{g}_{MN}\hat{g}^{SD}\hat{g}^{PK}\hat{\nabla}_{P}\hat{\nabla}_{S}{\varpi}_{KD}+\frac{1}{2}\hat{g}_{MN}\hat{g}^{SD}\hat{\square}{\varpi}_{SD}\nonumber
\\
&\qquad\qquad\qquad\qquad-2\,\hat{g}^{PK}\left(\hat{\nabla}_{N}{\varpi}_{MK}+\hat{\nabla}_{M}{\varpi}_{KN}-\hat{\nabla}_{K}{\varpi}_{MN}\right)A_{,P}+6\varpi_{MN}\hat{g}^{PS}\,A_{,S}A_{,P}\nonumber
\\
&\qquad\qquad\qquad\qquad\qquad\qquad+2\hat{g}_{MN}\hat{g}^{PK}\hat{g}^{SD}\left(\hat{\nabla}_{K}{\varpi}_{PD}+\hat{\nabla}_{P}{\varpi}_{KD}-\hat{\nabla}_{D}{\varpi}_{PK}\right)A_{,S}-4\,\varpi_{MN}\hat{\square}A\nonumber
\\
&
\qquad\qquad\qquad\qquad\qquad\qquad\qquad\qquad-6\hat{g}_{MN}\hat{g}^{PD}\hat{g}^{SJ}\varpi_{DJ}\,A_{,S}A_{,P}+4\,\hat{g}_{MN}\hat{g}^{PS}\hat{g}^{KD}\varpi_{SD}\hat{\nabla}_{P}\hat{\nabla}_{K}A,
\end{align}
where $\hat{g}_{MN}=e^{2A}g_{MN}$ and $\hat{\nabla}$ is a covariant derivative compatible with $\hat{\boldsymbol{g}}$. Introducing the assumed constraints, $\varpi_{Mi}=A_{,\mu}=0$, fixing the gauge $\hat{\nabla}^{M}\varpi_{MN}=\hat{g}^{MN}\varpi_{MN}=0$, noticing the commutation properties of the covariant derivatives\footnote{$\hat{g}^{PK}\hat{\nabla}_{P}\hat{\nabla}_{N}{\varpi}_{MK}=\hat{g}^{PK}\hat{\nabla}_{N}\hat{\nabla}_{P}{\varpi}_{MK}-\hat{g}^{SD}\hat{g}^{PK}\varpi_{SK}{\hat{R}}_{DMPN}+\hat{g}^{SD}\varpi_{MS}{\hat{R}}_{DN}$.} and considering that\footnote{Which is true whenever one assumes the gauge condition $\hat{g}^{MN}\varpi_{MN}=0$.} $\varpi_{\rho\kappa}\hat{R}^{\rho\kappa}=0$, one obtains the first order perturbed Einstein tensor in the form of
\begin{equation}
\delta G_{MN}=\frac{1}{2}\hat{g}^{\zeta \delta}\varpi_{\zeta \left(M\right.}{\hat{R}}_{\left.N\right)\delta }-\varpi^{\delta\rho}{\hat{R}}_{\delta M \rho N}-\frac{1}{2}\hat{\square}{\varpi}_{MN}+2\,\hat{\sigma}^{ij}A_{,i}\hat{\nabla}_{j}{\varpi}_{MN}+\varpi_{MN}\left(6\hat{\sigma}^{ij}\,A_{,i}A_{,j}-4\,\hat{\square}A-\frac{1}{2}\hat{R}\right),\label{perturbationeinstein}
\end{equation}
where $t_{\left(MN\right)}=t_{MN}+t_{NM}$.

One also notices that the metric fluctuations induce perturbations in the stress energy tensor. In particular, in the context of our analysis, only stress energy tensors constructed out of scalar fields shall be considered. One would also expect perturbations in the configuration of the scalar fields, but the scalar and tensorial perturbations are completely decoupled one from each other when linear perturbations are considered \cite{Csaki2000} (see Appendix \ref{decoupling} for further details). For this reason, when dealing with the dynamics of the gravitational field, one can disregard the perturbations of the scalar fields.

The stress energy tensor calculated out of metric $\bar{\boldsymbol{g}}$ is then given by
\begin{equation}
\bar{T}_{MN}={T}_{MN}-e^{-2A}\varpi_{MN}\left(\frac{{g}^{KS}\phi_{,S}\phi_{,K}}{2}+\frac{{g}^{KS}\zeta_{,S}\zeta_{,K}}{2}+V\right).
\end{equation}

Following the same stratagem from Ref.~\cite{Csaki2000}, one finds that
\begin{equation}
\varpi_{MP}{T^{P}}_{N}=-\varpi_{MN}\left(\frac{{g}^{KS}\phi_{,S}\phi_{,K}}{2}+\frac{{g}^{KS}\zeta_{,S}\zeta_{,K}}{2}+V\right),
\end{equation}
which can be used to derive the perturbations of the stress energy tensor,
\begin{equation}
\delta T_{MN}=4M^{4}\hat{g}^{\rho \zeta}\varpi_{M\rho}\hat{R}_{\zeta N}+4M^{4}\varpi_{MN}\left(6\hat{g}^{ij}A_{,i}A_{,j}-4\hat{\square}A-\frac{1}{2} \hat{R}\right).\label{perturbationstress}
\end{equation}

Finally, equalizing the first order contributions, $\delta G_{MN}$ and $\delta T_{MN}$, from Eqs.~\eqref{perturbationeinstein} and \eqref{perturbationstress}, through the Einstein equation, one finds
\begin{equation}\label{xxx}
-\frac{1}{2}\hat{\square}{\varpi}_{MN}+2\,\hat{\sigma}^{ij}A_{,i}\hat{\nabla}_{j}{\varpi}_{MN}=\frac{1}{2}\hat{g}^{\zeta \delta}\varpi_{\zeta \left[M\right.}{\hat{R}}_{\left.N\right]\delta }+\varpi^{\delta\rho}{\hat{R}}_{\delta M \rho N},
\end{equation}
where $t_{[MN]}=t_{MN}-t_{NM}$.
The above result be simplified by noticing that $\hat{\nabla}_{j}{\varpi}_{MN}={\varpi}_{MN,j}$, since $\hat{\Gamma}^{N}_{jM}=0$ and that $\hat{\boldsymbol{g}}$ is factorable. One thus has
\begin{equation}\label{xxxx}
\hat{\boldsymbol{R}}=\mathcal{R}^{\rho}_{\;\;\alpha\mu\nu}(x^{\kappa})\;\frac{\partial\;\;}{\partial x^{\rho}}\otimes\mathrm{d}x^{\alpha}\otimes\mathrm{d}x^{\nu}\otimes\mathrm{d}x^{\nu}+\hat{\Sigma}^{j}_{\;\;klc}(x^{s})\;\frac{\partial\;\;}{\partial x^{j}}\otimes\mathrm{d}x^{k}\otimes\mathrm{d}x^{l}\otimes\mathrm{d}x^{c},
\end{equation}
with $\mathcal{R}^{\rho}_{\;\;\alpha\mu\nu}$ and $\hat{\Sigma}^{j}_{\;\;klc}$ encoding the curvature of $\left(\mathbb{M}^{4},\boldsymbol{\omega}\right)$ and $\left(\mathbb{B}^{2},\boldsymbol{\hat{\sigma}}\right)$, respectively (cf. \cite{Gauy2022}). 
To summarize, one can still write the equation for the perturbation of the gravitational field for general bent branes,
\begin{equation}
\hat{\square}{\varpi}_{\mu\nu}-4\,\hat{\sigma}^{ij}A_{,i}{\varpi}_{\mu\nu,j}=
\omega^{\zeta \delta}\varpi_{\zeta \left[\nu\right.}{\mathcal{R}}_{\left.\mu\right]\delta }-2\varpi^{\delta\rho}{\mathcal{R}}_{\delta \mu \rho \nu},\label{bentperturbation}
\end{equation}
and notice that, for solutions \eqref{metricI}-\eqref{metricV}, only the Ricci tensor is written as $\mathcal{R}_{\mu\nu}=\Lambda \omega_{\mu\nu}$. For further simplifications, a maximally symmetric spacetime $\left(\mathbb{M}^{4},\boldsymbol{\omega}\right)$ can be assumed, i.e. ${\mathcal{R}}_{\delta\mu \rho \nu}=\left(\omega_{\delta\rho}\omega_{\mu\nu}-\omega_{\delta\nu}\omega_{\mu\rho}\right){\Lambda}/{3}$, such that Eq.~\eqref{bentperturbation} is thus simplified into the form of
 \begin{equation}
 \hat{\square}{\varpi}_{\mu\nu}-4\,\hat{\sigma}^{ij}A_{,i}{\varpi}_{\mu\nu,j}=\frac{2\Lambda}{3}\varpi_{\mu\nu},\label{symmetricperturbation}
 \end{equation}
which is essential in discussing intersecting branes. 

Finally, considering the perturbative approach applied to the action, one has
\begin{equation}\label{actiontotal}
	\delta S=2{M}^{4}\int\mathrm{d}^{6}xe^{-4A}\sqrt{-\hat{g}}\left[-\frac{1}{4}\hat{\nabla}^{K}\varpi^{\mu\nu}\hat{\nabla}_{K}\varpi_{\mu\nu}
	-\hat{g}^{ij}\varpi^{\mu\nu}\hat{\nabla}_{j}\varpi_{\mu\nu}A_{,i}+\varpi^{\mu\nu}\varpi_{\mu\nu}\left(2A^{,i}A_{,i}-\frac{\Lambda}{6}-\frac{1}{2}\hat{\square}A\right)\right],
\end{equation}
which will later select the localized solutions of Eq.~\eqref{symmetricperturbation}, which namely describe the four dimensional gravity.

\subsection{The Quantum Mechanical Analogy}
Eq.~\eqref{symmetricperturbation} can be readily refined by assuming a separation of variables algorithm,
\begin{equation}
\varpi_{\mu\nu}=\sum_{m\in\mathbb{I}}\Phi_{m}(u,v)\widetilde{\varpi}^{m}_{\mu\nu}(x^{\rho}),\label{separtionofvariables}
\end{equation}
which implies into two equations,
\begin{equation}
\left(\boldsymbol{\varDelta}-m^{2}\right)\widetilde{\varpi}^{m}_{\mu\nu}=0,\label{graviton}
\end{equation}
and
\begin{equation}\label{locality}
{\hat{\triangle}}^{2}\Phi_{m}-4\hat{\sigma}^{ij}A_{,i}\Phi_{m,j}=\left(\frac{2\Lambda}{3}-m^{2}\right)\Phi_{m},
\end{equation}
where $\boldsymbol{\varDelta}=\omega^{\alpha\beta}\hat{\nabla}_{\alpha}\hat{\nabla}_{\beta}$, $\hat{\triangle}^{2}=\hat{\sigma}^{ij}\hat{\nabla}_{i}\hat{\nabla}_{j}$ and $m^{2}$ is a separation constant.

A simpler form of Eq.~\eqref{locality} can be achieved by rescaling the field $\Phi_{m}=e^{2A}\chi_{m}$, so as to give
\begin{equation}\label{schroedinger}
	-{\hat{\triangle}}^{2}\chi_{m}+2\left(2\hat{\sigma}^{ij}A_{,i}A_{,j}-{\hat{\triangle}}^{2}A\right)\chi_{m}=\left(m^{2}-\frac{2\Lambda}{3}\right)\chi_{m},
\end{equation}
which shall be identified with a time independent Schr\"odinger-like equation in curved space $\left(\mathbb{B}^{2},\hat{\boldsymbol{\sigma}}\right)$, with the energy $E_{QM}=m^{2}-{2\Lambda}/{3}$ and the ``quantum mechanical'' potential
\begin{equation}
V_{QM}(u,v)=2\left(2A^{,i}A_{,i}-{\hat{\triangle}}^{2}A\right)=2\left[2\hat{\sigma}^{ij}A_{,i}A_{,j}-\frac{1}{\sqrt{\hat{\sigma}}}\left(\sqrt{\hat{\sigma}}\hat{\sigma}^{ij}A_{,i}\right)_{,j}\right].
\end{equation}

A ``flat'' Schr\"odinger-like equation is indeed solely justified for a conformally flat metric $(\bar{f}=\bar{h}=0\iff\hat{\boldsymbol{\sigma}}=\boldsymbol{\gamma})$, with $\gamma=\text{diag}\left({1,1}\right)$. Otherwise the curvature of $\left(\mathbb{B}^{2},\hat{\boldsymbol{\sigma}}\right)$ is completely arbitrary, and the curvature intricacies introduce the possibility of some additional localization aspect.

Following the same algorithm from Eq.~\eqref{separtionofvariables}, and rescaling the scalar field by $\Phi_{m}=e^{2A}\chi_{m}$, the action for the perturbations can be written as
\begin{equation}
\delta S=-\frac{{M}^{4}}{2}\sum_{m_{1},m_{2}\in\mathbb{I}}\int\mathrm{d}^{2}y\sqrt{\hat{\sigma}}\chi_{m_{2}}\chi_{m_{1}}\int\mathrm{d}^{4}x\sqrt{-{\omega}}\left(\hat{\nabla}^{\kappa}\widetilde{\varpi}_{m_{1}}^{\mu\nu}\hat{\nabla}_{\kappa}\widetilde{\varpi}^{m_{2}}_{\mu\nu}+{m_{1}}^{2}\widetilde{\varpi}_{m_{1}}^{\mu\nu}\widetilde{\varpi}^{m_{2}}_{\mu\nu}\right),
\end{equation}
which corresponds to the same problem driven by Eq.~\eqref{graviton}, if and only if the gravitational modes are
\begin{enumerate}
	\item Normalizable, i.e. $0<\displaystyle\int\mathrm{d}^{2}y\sqrt{\hat{\sigma}}\chi_{m}{}^{2}\neq\infty,\;\;\forall m \in \mathbb{I}$;
	\item Orthogonal, i.e. $\displaystyle\int\mathrm{d}^{2}y\sqrt{\hat{\sigma}}\chi_{m_{2}}\chi_{m_{1}}=0\text{ if }m_{1}\neq m_{2}$.
\end{enumerate}

When the above conditions are satisfied, for operators compatible with $\boldsymbol{\omega}$, the action reads
\begin{equation}\label{reducedaction}
\delta S=-\frac{{M}^{4}}{2}\sum_{m\in\mathbb{I}}\int\mathrm{d}^{2}y\sqrt{\hat{\sigma}} {\chi_{m}}^{2}\int\mathrm{d}^{4}x\sqrt{-{\omega}}\left(\varDelta^{\kappa}\widetilde{\varpi}_{m}^{\mu\nu}\varDelta_{\kappa}\widetilde{\varpi}^{m}_{\mu\nu}+{m}^{2}\widetilde{\varpi}_{m}^{\mu\nu}\widetilde{\varpi}^{m}_{\mu\nu}\right),
\end{equation}
which does not provide a coupling between the different states $\widetilde{\varpi}_{m_{i}}$. In this scope, variations of action \eqref{reducedaction} lead to the Eq.~\eqref{graviton}. In particular, in higher dimensional theories there should be as many massive gravitons as can be fit in the set $\mathbb{I}$, which contains the normalizable states of the Schr\"odinger equation \eqref{schroedinger}. 

The requirement of normalizability again ratifies the quantum mechanical analogy. The gravitational modes $\chi_{m}$ must satisfy a Schr\"odinger-like equation and be normalized in the curved space $(\mathbb{B}^{2},\hat{\boldsymbol{\sigma}})$, and the localization of gravity at the vicinity of the brane now becomes contingent on the ``quantum mechanical'' problem described by Eqs.~\eqref{schroedinger} and \eqref{reducedaction}.
The problem of locality is then reduced to solving the ``quantum mechanical'' problem described by Eq.~\eqref{schroedinger} according to the normalization condition from Eq.~\eqref{reducedaction}.

	\subsection{The Newtonian Limit}\label{newtonconstant}
	
	Rigorously, for thick braneworlds, matter fields should be smeared over the bulk. But, for the sake of simplicity, the gravitational potential will be generated by a point-like source of mass\footnote{This same strategy was employed by Ref.~\cite{Csaki2000}.} $\mathcal{M}$. The action of a point-like particle, up to the first order in $\varpi$, is written as (see Appendix \ref{ParticleStress} for further details)
	\begin{equation}
	\delta S_{p}=-\frac{\mathcal{M}}{2}\sum_{m\in\mathbb{I}}\int \mathrm{d}^{6}x \;\chi_{m}\delta(x^{Q}-x^{P})\widetilde{\varpi}^{m}_{\mu\nu}\mathsf{v}^{\mu}\mathsf{v}^{\nu}.\label{particlefirst}
	\end{equation}
where $\mathsf{v}$ represents the velocity of the particle in space-time and $x^{P}$ is the position in $\mathbb{E}^{6}$. After a re-parameterization of the proper time so to satisfy $\omega_{\mu\nu}\mathsf{v}^{\mu}\mathsf{v}^{\nu}=-1$, for a particle at $y^{i}_{0}$, the total action, to the leading order in $\varpi$, becomes
	\begin{align}
	\delta S=&-\sum_{m\in\mathbb{I}}\left[\frac{{M}^{4}}{2}\int\mathrm{d}^{2}y\,\sqrt{\hat{\sigma}}\, {\chi_{m}}^{2}\int\mathrm{d}^{4}x\sqrt{-{\omega}}\left(\varDelta^{\kappa}\widetilde{\varpi}_{m}^{\mu\nu}\varDelta_{\kappa}\widetilde{\varpi}^{m}_{\mu\nu}+{m}^{2}\widetilde{\varpi}_{m}^{\mu\nu}\widetilde{\varpi}^{m}_{\mu\nu}\right)\nonumber
	\right.\\&\qquad\qquad\qquad\qquad\qquad\qquad\qquad\qquad\qquad\qquad\qquad\qquad\left.
	+\frac{\mathcal{M}}{2}e^{2A(y_{0}^{i})}\chi_{m}(y_{0}^{i})\int\mathrm{d}x^{4}\delta(x^{\mu}-x^{\alpha})\widetilde{\varpi}^{m}_{\mu\nu}\mathsf{v}^{\mu}\mathsf{v}^{\nu}\right],
	\end{align}
	where $x^{\alpha}$ and $y_{0}^{i}$ represent the position of the point-like particle in $\mathbb{M}^{4}$ and $\mathbb{B}^{2}$, respectively. Varying with respect to $\varpi$ leads to the equations of motion
	\begin{equation}\label{xxxxx}
	\left(\varDelta^{2}-m^{2}\right)\tilde{\varpi}^{m}_{\mu\nu}=\frac{\mathcal{M}e^{2A(y_{0}^{i})}\chi_{m}(y_{0}^{i})}{2M^{4}\displaystyle\int\mathrm{d}^{2}y\,\sqrt{\hat{\sigma}}\,{\chi_{m}}^{2}}\frac{\mathsf{v}_{\mu}\mathsf{v}_{\nu}\delta(x^{\mu}-x^{\alpha})}{\sqrt{-\omega}}.
	\end{equation}
	
	Hence, the Newtonian limit is obtained through the $00$ component of Eq.~\eqref{xxxxx}, with $v^{1}=v^{2}=v^{3}=0$ and $v^{0}=1/\sqrt{-\omega_{00}}$, for a static particle. Also supposing that the particle finds itself at the center of a system of coordinates ($\mathsf{r}=0$), the configuration has been stabilized, {\em i.e.} with $\widetilde{\varpi}$ independent of time, and the space displacements for identifying the Newtonian potential are not of cosmological scale ($\mathsf{r}\sim1/\sqrt{\Lambda}$). It allows for setting the approximation $\boldsymbol{\omega}\approx\boldsymbol{\eta}$ so to simplify the equations of motion into the form of
	\begin{equation}
	\left(\nabla^{2}-m^{2}\right)\tilde{\varpi}^{m}_{00}=\frac{\mathcal{M}e^{2A(y_{0}^{i})} \chi_{m}(y_{0}^{i})}{2M^{4}\displaystyle\int\mathrm{d}^{2}y\sqrt{\hat{\sigma}}{\chi_{m}}^{2}}\delta(\boldsymbol{\mathsf{r}}),
	\end{equation}
	with the direct solution\footnote{The solution is true for as long as $m\in\mathbb{R}$, otherwise, i.e. $m\in\mathbb{C}$ ($Re(m)=0$), one would find $\frac{\cos\left(m \mathsf{r}\right)}{\mathsf{r}}$ instead of $\frac{e^{-m\mathsf{r}}}{ \mathsf{r}}$.}
	\begin{equation}
	\tilde{\varpi}^{m}_{00}=\frac{\mathcal{M}e^{2A(y_{0}^{i})} \chi_{m}(y_{0}^{i})}{8\pi M^{4}\displaystyle\int\mathrm{d}^{2}y\sqrt{\hat{\sigma}}{\chi_{m}}^{2}}\frac{e^{-m\mathsf{r}}}{ \mathsf{r}},
	\end{equation}
	so to obtain the Newtonian potential,
	\begin{equation}
	\phi_{N}(\mathsf{r})=\frac{\varpi_{00}(y_{0}^{i},\mathsf{r})}{4}=\sum_{m\in\mathbb{I}}\frac{\mathcal{M}e^{4A(y_{0}^{i})}\left[\chi_{m}(y_{0}^{i})\right]^{2}}{32\pi M^{4}\displaystyle\int\mathrm{d}^{2}y\sqrt{\hat{\sigma}}{\chi_{m}}^{2}}\frac{e^{-m\mathsf{r}}}{ \mathsf{r}}\label{newtonianpotential}
	\end{equation}
which can be considered in order to make the connection with the phenomenology of braneworlds.

\section{The Quantum Analogue Problem For Intersecting Branes}\label{quantum}

The prescription of the preceding subsections can be readily particularized to intersecting thick braneworlds. The quantum problem, Eq.~\eqref{schroedinger}, for the geometries described by metric \eqref{intersectingmetric} is characterized by the equation
\begin{align}
&-e^{\left(p-1\right)\tilde{A}+\tilde{f}}\left[e^{\left(p-1\right)\tilde{A}+\tilde{f}}\chi_{m,u}\right]_{,u}-e^{-p\hat{A}+\hat{h}}\left(e^{-p\hat{A}+\hat{h}}\chi_{m,v}\right)_{,v}+2\Bigg[2e^{2\left(p-1\right)\tilde{A}+2\tilde{f}}\tilde{A}_{,u}{}^{2}+2e^{-2p\hat{A}+2\hat{h}}\hat{A}_{,v}{}^{2}\Bigg.\nonumber
\\
&\qquad\qquad\qquad\Bigg.-e^{\left(p-1\right)\tilde{A}+\tilde{f}}\left[e^{\left(p-1\right)\tilde{A}+\tilde{f}}\tilde{A}_{,u}\right]_{,u}-e^{-p\hat{A}+\hat{h}}\left(e^{-p\hat{A}+\hat{h}}\hat{A}_{,v}\right)_{,v}\Bigg]\chi_{m}=e^{2p\tilde{A}}e^{-2\left(p-1\right)\hat{A}}\left(m^{2}-\frac{2\Lambda}{3}\right)\chi_{m}.\label{interschroedinger}
\end{align}
A separation of variables technique, with $\chi_{m}=\sum_{k\in\mathbb{K}}\tilde{\chi}_{k}(u)\hat{\chi}_{km}(v)$, can be applied to Eq.~\eqref{interschroedinger} if:
\begin{enumerate}
	\item $m=\sqrt{2\Lambda/3}$ (i.e. the zero mode);
	\item $p=0$ (or $p=1$) (i.e. for models $III$, $IV$ and $V$).
\end{enumerate}

If $p\neq0$ the solutions of Eq.~\eqref{interschroedinger} are not separable unless $m=\sqrt{2\Lambda/3}$, which can always be expressed by $\chi_{0}=be^{-2A}$ (see \cite{Csaki2000}). On the other hand, whenever $p=0$ the variables are separable and Eq.~\eqref{interschroedinger} is reduced to two equations
\begin{equation}
-e^{\tilde{f}-\tilde{A}}\left[e^{\tilde{f}-\tilde{A}}\tilde{\chi}_{k,u}\right]_{,u}+\left[4e^{2\tilde{f}-2\tilde{A}}\tilde{A}_{,u}{}^{2}-2e^{\tilde{f}-\tilde{A}}\left(e^{\tilde{f}-\tilde{A}}\tilde{A}_{,u}\right)_{,u}+k\right]\tilde{\chi}_{k}=0,\label{schroedingeru}
\end{equation}
and
\begin{equation}
-e^{\hat{h}}\left(e^{\hat{h}}\hat{\chi}_{mk,v}\right)_{,v}+\left[4e^{2\hat{h}}\hat{A}_{,v}{}^{2}-2e^{\hat{h}}\left(e^{\hat{h}}\hat{A}_{,v}\right)_{,v}-k\right]\hat{\chi}_{mk}=e^{2\hat{A}}\left(m^{2}-\frac{2\Lambda}{3}\right)\hat{\chi}_{mk}.\label{schroedingerv}
\end{equation}

The normalization of the gravitational modes, for the $p=0$ models, is fixed by
\begin{equation}\label{normalization}
\int \sqrt{\hat{\sigma}}\chi_{m}{}^{2} \mathrm{d}u\mathrm{d}v=\sum_{k\in\mathbb{K}}\int \left(e^{\tilde{A}-\tilde{f}}\tilde{\chi}_{k}{}^{2}\right)\mathrm{d}u\,\int\left( e^{2\hat{A}-\hat{h}}\hat{\chi}_{mk}{}^{2} \right)\mathrm{d}v<\infty.
\end{equation}

Eq.~\eqref{schroedingeru} and the normalization condition for $\tilde{\chi}_{k}$ are precisely the same as in five dimensions\footnote{Since coordinates can always be chosen such that $\tilde{f}=\tilde{A}$.}. However, Eq.~\eqref{schroedingerv} represents a distinct problem, as it shall be discussed in the following. 

\subsection{Models $I$ and $II$}\label{model I and II}

As argued before, for the $p\neq0$ constructions, i.e. models $I$ and $II$, one can merely determine the zero modes \cite{Gauy2022} as, respectively,
\begin{equation}
\chi^{I}_{0}=B^{I}\sqrt{\cosh \left(2c_{u}u\right)\left|\cos \left(2c_{v}v\right)\right|},\label{waveI}
\end{equation}
and
\begin{equation}
\chi^{II}_{0}=B^{II}\sqrt{\left|\cos \left(2c_{u} u\right)\cos \left(2c_{v}v\right)\right|},\label{waveII}
\end{equation}
where $B^{I}$ and $B^{II}$ are normalization constants. Wave functions \eqref{waveI} and \eqref{waveII} are normalizable, since
\begin{equation}
\int \sqrt{\hat{\sigma}}\left(\chi^{I}_{0}\right)^{2} \mathrm{d}u\mathrm{d}v=\frac{\left({B^{I}}\right)^{2}}{4c_{u}c_{v}}\int^{\infty}_{-\infty}\operatorname{sech}^{\frac{p-3}{4}}\left(x\right)\mathrm{d}x\int^{\frac{\pi}{2}}_{-\frac{\pi}{2}}\cos^{\frac{p+2}{4}}\left(\varphi\right)\mathrm{d}\varphi=\frac{\left({B^{I}}\right)^{2}}{4c_{u}c_{v}}\frac{\pi  \Gamma \left(\frac{p-3}{8}\right) \Gamma \left(\frac{p+6}{8}\right)}{\Gamma \left(\frac{p+1}{8}\right) \Gamma \left(\frac{p+10}{8}\right)},\label{normalization1}
\end{equation}
and
\begin{equation}
\int \sqrt{\hat{\sigma}}\left(\chi^{II}_{0}\right)^{2} \mathrm{d}u\mathrm{d}v=\frac{\left({B^{II}}\right)^{2}}{4c_{u}c_{v}}\int^{\frac{\pi}{2}}_{-\frac{\pi}{2}}\sec^{\frac{p-3}{4}}\left(\theta\right)\mathrm{d}\theta\int^{\frac{\pi}{2}}_{-\frac{\pi}{2}}\cos^{\frac{p+2}{4}}\left(\varphi\right)\mathrm{d}\varphi=\frac{\left({B^{II}}\right)^{2}}{4c_{u}c_{v}}\frac{\pi  \Gamma \left(\frac{7-p}{8}\right) \Gamma \left(\frac{p+6}{8}\right)}{\Gamma \left(\frac{11-p}{8}\right) \Gamma \left(\frac{p+10}{8}\right)}.\label{normalization2}
\end{equation}

\begin{figure}[!htb]
	\subfloat[]{\includegraphics[scale=0.63]{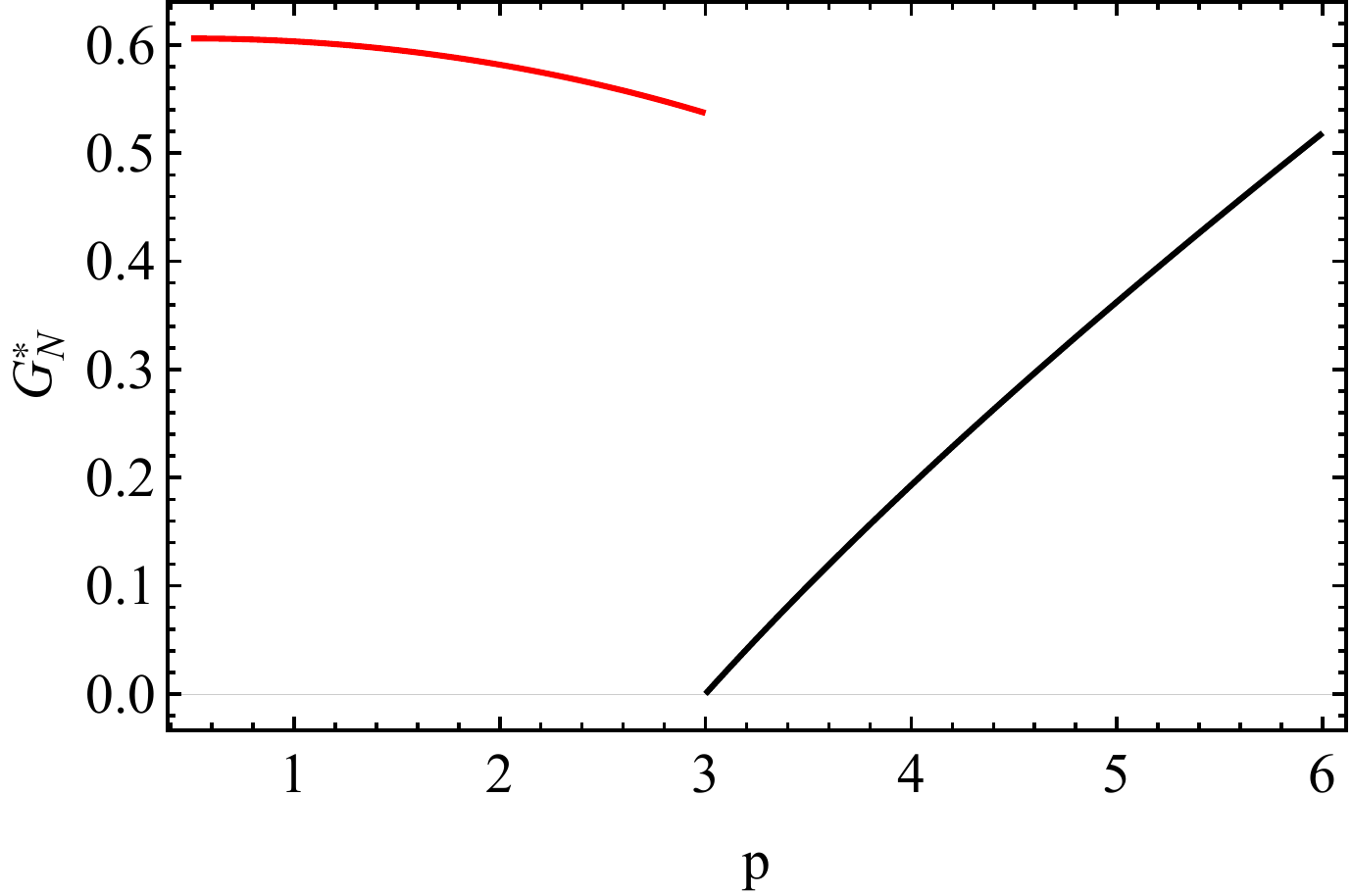}}$\;\;$
	\subfloat[]{\includegraphics[scale=0.63]{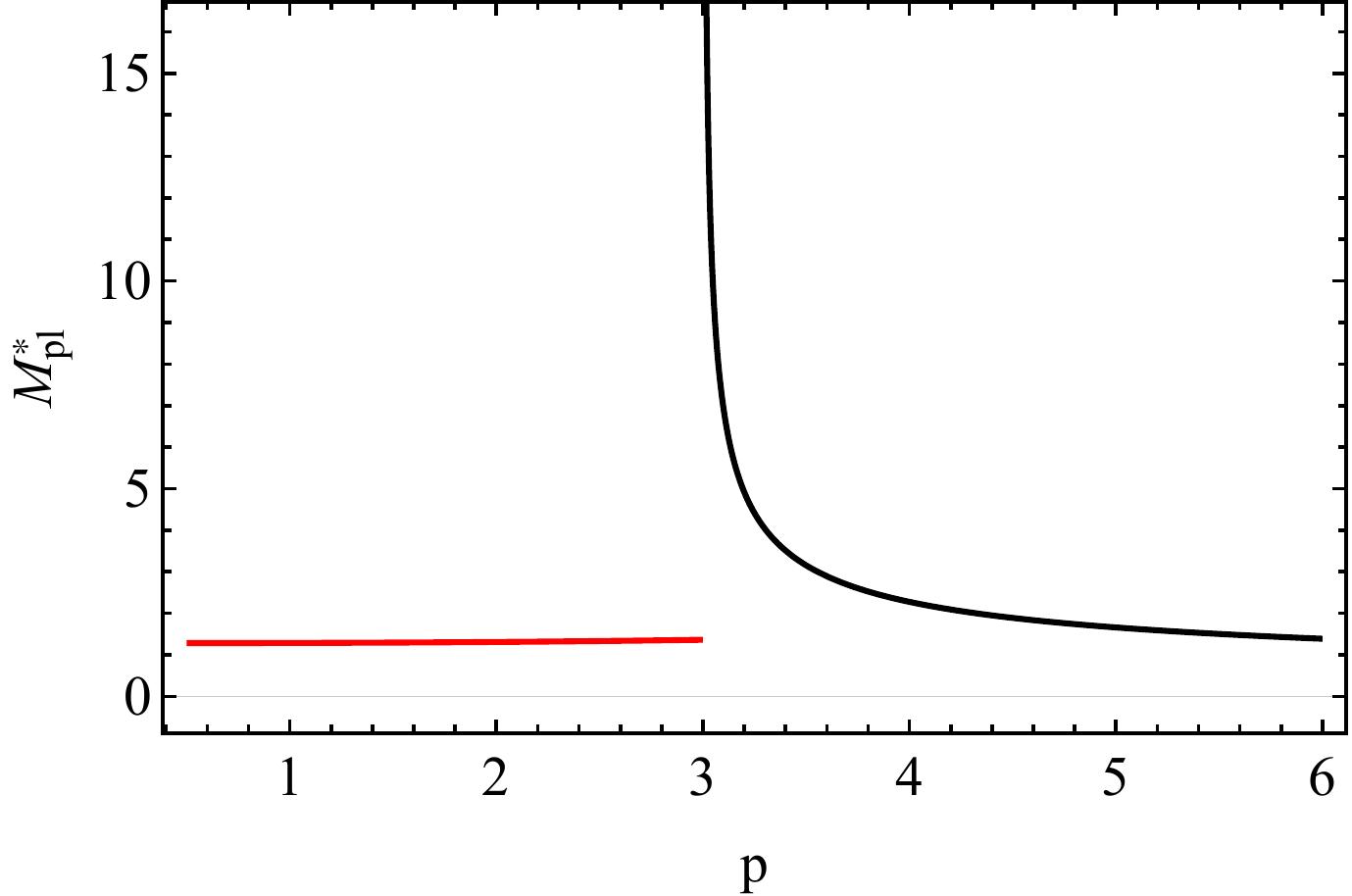}}
	\caption{(Color online) (a) The gravitational constant $G^{*}_{N}=2^{3}\pi^{2} M^{4} G_{N}/c_{u}c_{v}$ as a function of $p$. (b) The Planck Scale $M^{*}_{pl}=2M_{pl}\sqrt{c_{u}c_{v}}/M^{2}\sqrt{\pi}$ as a function of $p$. The plots are for model $I$ (black line) and $II$ (red line).}\label{figure1}
\end{figure}

Additional massive modes are not straightforwardly obtained, thus the corrections to the Newtonian gravity cannot be evaluated. Regardless, the Newtonian gravitational constant can be settled from Eqs. \eqref{normalization1} and \eqref{normalization2}:
\begin{equation}
G^{I}_{N}=\frac{1}{2^{5}\pi \left(M^{I}_{pl}\right)^{2}}=\frac{c_{u}c_{v}\Gamma \left(\frac{p+1}{8}\right) \Gamma \left(\frac{p+10}{8}\right)}{2^{3}\pi^{2} M^{4}\Gamma \left(\frac{p-3}{8}\right) \Gamma \left(\frac{p+6}{8}\right)},
\end{equation}
and
\begin{equation}
G^{II}_{N}=\frac{1}{2^{5}\pi \left(M^{II}_{pl}\right)^{2}}=\frac{c_{u}c_{v}\Gamma \left(\frac{11-p}{8}\right) \Gamma \left(\frac{p+10}{8}\right)}{2^{3}\pi^{2} M^{4}\Gamma \left(\frac{7-p}{8}\right) \Gamma \left(\frac{p+6}{8}\right)}.
\end{equation}
The gravitational constant $G_{N}$ and the associated Planck scale $M_{pl}$ are depicted in Fig. \ref{figure1}.

Ordinarily, for brane models, the gravitational scale of model $II$ is controlled by the parameters $c_u$ and $c_v$. In contrast, the strength of gravity for model $I$ can be set by choosing the parameter $p$, which can take widely different values when $p \gtrsim 3$. This is relevant when addressing the hierarchy problem, because a large Planck scale can be achieved for any value of $c_{u}$ and $c_{v}$. Unfortunately, model $I$ is plagued by several modeling issues: the stress energy tensor has singularities and the total defect energy formation is infinite \cite{Gauy2022}.

	Even though models $I$ and $II$ are constructed from the same solution \cite{Gauy2022}, the transition from one to the other involves redefining $c_u$ from completely real, for model $I$, to completely imaginary, for model $II$. This transition will imply in a discontinuous behavior for any physical constant, as exhibited by the discontinuity depicted in Fig. \ref{figure1} at the boundary between models $I$ and $II$.

\subsection{Model $III$}
Eq.~\eqref{schroedingerv} for model $III$ (represented by metric \eqref{metricIII}) is expressed by
\begin{equation}
-\hat{\chi}^{III}_{mk,vv}-\left[C+k+C \sec ^2\left(2 \sqrt{C} v\right)\right]\hat{\chi}^{III}_{mk}=\displaystyle\frac{m^{2}\hat{\chi}^{III}_{mk}}{\sqrt{\cos\left(2\sqrt{C}v\right)}}.\label{schoerodingerIII}
\end{equation}

\begin{figure}[!htb]
		\includegraphics[scale=0.63]{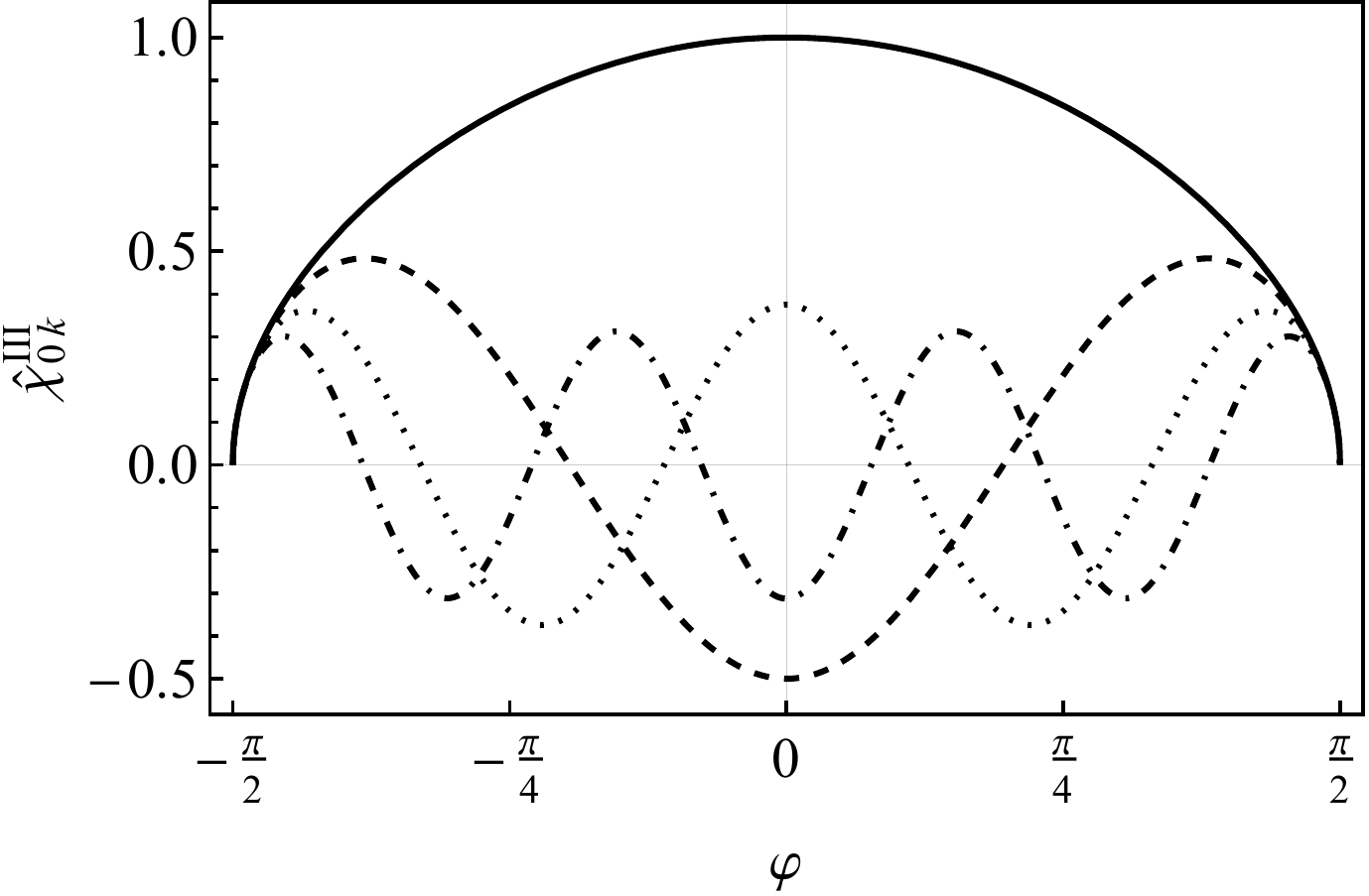}
		\caption{(Color online) The degeneracies of the zero mode of model $III$ as functions of $\varphi=2\sqrt{C}v$ for $j=0$ (full line), $j=1$ (dashed line), $j=2$ (dotted line) and $j=3$ (dot-dashed line).}\label{figure2}
	\end{figure}

For the general massive case, Eq.~\eqref{schoerodingerIII} cannot be straightforwardly integrated, and the unique immediate solution is the zero mode, which is proportional to the warp factor when $k=0$. For any $k$, the zero mode can be written as

	\begin{equation}
	\hat{\chi}^{III}_{0k}= B^{III}_{0k} \sqrt{\cos \left(2 \sqrt{C} v\right)} \, _2F_1\left[\frac{1}{4}\left(1-\sqrt{1+\frac{k}{C}}\right),\frac{1}{4}\left(1+\sqrt{1+\frac{k}{C}}\right);1;\cos ^2\left(2 \sqrt{C} v\right)\right],\label{stateIII}
	\end{equation}
As it will be further argued in Subsec. \ref{model V}, the allowed wave functions must be continuously differentiable. The first derivative of \eqref{stateIII} is necessarily discontinuous at $v=0$, unless the first or second argument of the hypergeometric function, $_2F_1$, is a non-positive integer $-j$, $j\in\mathbb{N}$, or $k=8 C j (2 j+1)$. The degeneracies of the zero mode, for model $III$, are depicted in Fig. \ref{figure2}.

Eq.~\eqref{schoerodingerIII} can be still more simplified if one makes $C=0$ and $v=r\varphi$, leading to a trivial extension of five dimensional models \cite{Gauy2022},
	\begin{equation}
	-\hat{\chi}^{III}_{mk,\varphi\varphi}=r^{2}\left(m^{2}+k\right)\hat{\chi}_{mk},
	\end{equation}
which results into
	\begin{equation}
	\hat{\chi}^{III}_{mk}=B \cos\left(n\varphi\right),
	\end{equation}
where $m^{2}+k=n^{2}/r^{2}$, $B$ is a normalization constant and one assumes boundary conditions\footnote{With the intent of discarding the $\sin\left(n\varphi\right)$ terms.} such that $\hat{\chi}^{III}_{mk}=0$ at $\varphi=\pm \pi$. This trivial solution will be important for the sphere models, where one will be able to compare trivial constructions with a more sophisticated configuration, namely that one engendered by model $IV$.
	
	\subsection{Model $IV$}\label{modelIV}
	To simplify the analysis that follows, it is more convenient to work with coordinates such that 
	\begin{equation}
	v=\frac{2}{\sqrt{C}}\arcsin\left[\tanh\left(\frac{\sqrt{C}y}{2}\right)\right],\label{conformalcoordinates}
	\end{equation}
	which, from Eq.~\eqref{metricIV}, implies into the metric $IV$ recasted as
	\begin{equation}
	g^{IV}=\frac{4\Lambda}{3C}\operatorname{sech}^{2}\left(\frac{\sqrt{C}y}{2}\right)e^{-2\tilde{A}}\omega_{\mu\nu}\mathrm{d}x^{\mu}\mathrm{d}x^{\nu}+e^{-2\tilde{A}}\mathrm{d}u^{2}+e^{-2\tilde{A}}\operatorname{sech}^{2}\left(\frac{\sqrt{C}y}{2}\right)\mathrm{d}y^{2}.
	\end{equation}
After a rescaling of the wave function, $\hat{\chi}^{IV}_{mk}=\operatorname{sech}^{1/2}\left({\sqrt{C}y}/{2}\right)\hat{\xi}^{IV}_{mk}$, the Schr\"odinger-like equation \eqref{schroedingerv} becomes a P\"oschl-Teller equation \cite{Fluegge1998},
	\begin{equation}
	-\hat{\xi}^{IV}_{mk,yy}-\lambda\left(\lambda-1\right)\operatorname{sech}^2\left(z\right)\hat{\xi}^{IV}_{mk}=E\;\hat{\xi}^{IV}_{mk},\label{posch-teller}
	\end{equation}
	where $\lambda=2\sqrt{1+\frac{k}{C}}+\frac{1}{2}$, $E=\frac{3m^{2}}{\Lambda}-\frac{17}{4}$ and $z={\sqrt{C}y}/{2}$. The general solution of Eq. \eqref{posch-teller} is 
	\begin{equation}
	\hat{\xi}^{IV}_{mk}= B_1 \cosh^{\lambda}\left(z\right){}_2F_{1}\left[\alpha,\beta;\frac{1}{2};-\sinh^{2}\left(z\right)\right]+B_2 \cosh^{\lambda}\left(z\right)\sinh\left(z\right){}_2F_{1}\left[\alpha+\frac{1}{2},\beta+\frac{1}{2};\frac{3}{2};-\sinh^{2}\left(z\right)\right],\label{waveIV}
	\end{equation}
	with the parameters $\alpha=\frac{1}{2}\left(\lambda-\sqrt{-E}\right)$ and $\beta=\frac{1}{2}\left(\lambda+\sqrt{-E}\right)$. If $\lambda\leq1$ (i.e. $k\leq-15C/16$) or $E>0$ (i.e. $m^{2}>\frac{17}{12}\Lambda$) the above solutions correspond to propagating modes. Otherwise, imposing unitary boundary conditions \cite{Gremm2000,HERRERAAGUILAR2010} suppresses all propagating modes. In this case, the unitary spectrum shall be composed uniquely of bound states\footnote{Because they do not generate any flux into the singularities at the boundaries of space.}.
	
\begin{figure}[!htb]
		\includegraphics[scale=0.63]{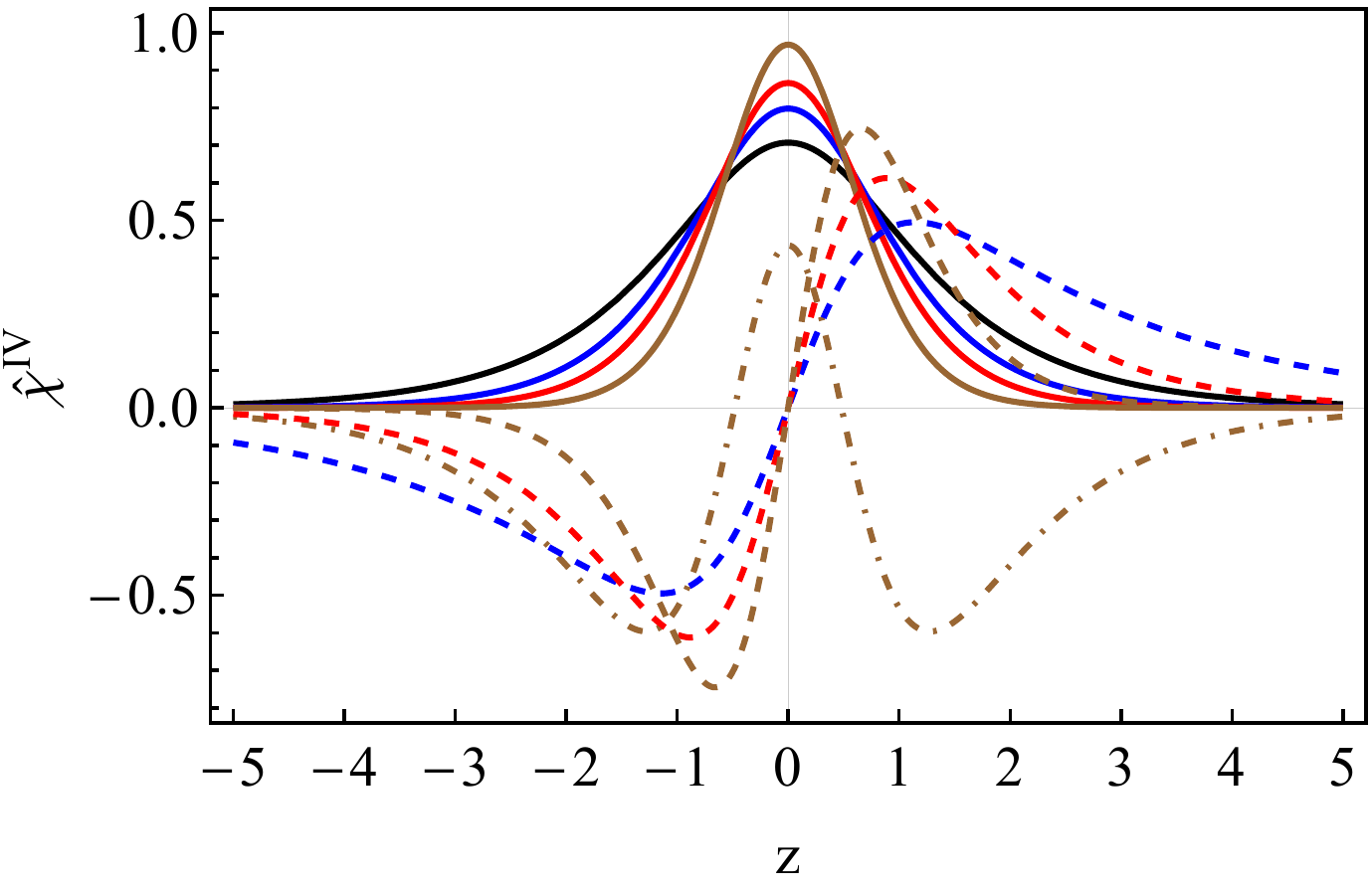}
		\caption{(Color online) Normalized gravitational wave functions $\hat{\chi}^{IV}$ for model $IV$ as functions of $z=v\sqrt{C}/2$, for $k/C=-7/16$ (black lines), $k/C=0$ (blue lines), $k/C=9/16$ (red lines) and $k/C=33/16$ (brown lines). The full, dashed and dot-dashed lines represent $j=0$, $l=0$ and $j=1$, respectively.}\label{modes_IV}
	\end{figure}

	The normalization condition for $\hat{\xi}_{mk}$, in terms of conformal coordinates \eqref{conformalcoordinates}, reads 
	\begin{equation}
	0<\int^{\infty}_{-\infty} \left(\hat{\xi}^{IV}_{mk}\right)^{2}\mathrm{d}y\neq\infty.
	\end{equation}
	Therefore, normalizable modes will exist for $\lambda>1$ and negative $E$, where
	\begin{equation}
	\hat{\xi}^{IV+}_{jk}=B^{+}_{jk} \cosh^{2\sqrt{1+\frac{k}{C}}+\frac{1}{2}}\left(z\right){}_2F_{1}\left[\frac{1}{2}+j,2\sqrt{1+\frac{k}{C}}-j;\frac{1}{2};-\sinh^{2}\left(z\right)\right]
	\end{equation}
	represents normalizable even eigenstates, with mass eigenvalues
	\begin{equation}
	m^{IV+}_{jk}=\sqrt{\frac{\Lambda}{3}\left[\frac{17}{4}-\left(2\sqrt{1+\frac{k}{C}}-\frac{1}{2}-2j\right)^{2}\right]},\;\;0\leq j< \sqrt{1+\frac{k}{C}}-\frac{1}{4};\label{massesIV+}
	\end{equation}
	and
	\begin{equation}
	\hat{\xi}^{IV-}_{lk}=B^{-}_{lk} \cosh^{2\sqrt{1+\frac{k}{C}}+\frac{1}{2}}\left(z\right)\sinh\left(z\right){}_2F_{1}\left[\frac{3}{2}+l,2\sqrt{1+\frac{k}{C}}-l;\frac{3}{2};-\sinh^{2}\left(z\right)\right],
	\end{equation}
	represents normalizable odd eigenstates, with mass eigenvalues
	\begin{equation}
	m^{IV-}_{lk}=\sqrt{\frac{\Lambda}{3}\left[\frac{17}{4}-\left(2\sqrt{1+\frac{k}{C}}-\frac{3}{2}-2l\right)^{2}\right]},\;\;0\leq l < \sqrt{1+\frac{k}{C}}-\frac{3}{4};\label{massesIV-}
	\end{equation}
	where $j$ and $l$ are natural numbers, and $B^{\pm}_{jk}$ are normalization constants. Some normalizable modes are depicted in Fig. \ref{modes_IV}. The masses, as described by Eqs. \eqref{massesIV+} and \eqref{massesIV-}, are strictly real valued as long as $-15C/16< k\leq 0$, while for positive values of $k$ some tachyonic states are allowed. 

	Interestingly, model $IV$ does not have a singular Ricci scalar\footnote{The singularities at the edges of space are associated with the Kretschmann scalar.}, but it exhibits a mass gap between the zero and massive modes. This is noteworthy because one would expect that a mass gap would imply in singularities of the Ricci scalar, which always happens for five dimensional models \cite{HERRERAAGUILAR2010}. Thus, one may conjecture that mass gaps and naked singularities are connected, but the singularity might be related with scalar invariants other than the Ricci scalar.

	\subsection{Model $V$}\label{model V}
	Finally, once applied to model $V$, Eq.~\eqref{schroedingerv} is cast into the form of
	\begin{equation}
	-\hat{\chi}^{V}_{mk,vv}-\sqrt{\frac{\left|\Lambda\right| }{3}}\tan \left(\sqrt{3\left|\Lambda\right|} v\right)\hat{\chi}^{V}_{mk,v}-\left[\frac{4\left|\Lambda\right|\sec^{2} \left(\sqrt{3\left|\Lambda\right|} v\right)}{3}+k\right]\hat{\chi}^{V}_{mk}=m^{2}\hat{\chi}^{V}_{mk}.\label{schoerodingerV}
	\end{equation}
	
	Rescaling $\hat{\chi}^{V}_{mk}$ by $\cos^{1/6}\left(\sqrt{3\left|\Lambda\right|} v\right)$ transforms the above result from Eq.~\eqref{schoerodingerV} into a trigonometric Pöschl-Teller equation, with the straightforward solution identified by
	\begin{equation}
	\hat{\chi}^{V}_{l}= B^{V}_{l}\cos ^{\frac{2}{3}}\left(\sqrt{3\left|\Lambda\right|}v\right) \, _2F_1\left[l,\frac{1}{2}-l;1;\cos ^2\left(\sqrt{3\left|\Lambda\right|}v\right)\right],\label{stateV}
	\end{equation}
with the eigenvalues identified by
	\begin{equation}
	m^{V}_{kl}= \sqrt{\left[12 l\left(l-\frac{1}{2}\right)-\frac{2}{3}\right] | \Lambda | -k},\label{massV}
	\end{equation}
where $B^{V}_{l}$ is a normalization constant.

The hypergeometric component, ${}_{2}F_{1}\left(\alpha,\beta;\gamma;z\right)$, have arguments such that $\gamma-\alpha-\beta=1/2$, thus a discontinuity of the first derivatives, at $v=0$, is only avoidable if either $\alpha$ or $\beta$ is a non-positive integer $-j$, $j\in \mathbb{N}$ (see Appendix \ref{Discontinuities} for further details). Otherwise, an unphysical discontinuity is identified for the stress energy tensor of the perturbations, since the latter depends on the first derivatives of \eqref{stateV}. Therefore, the allowed values of $l$ are given by $l=-j$ or $l=j+{1}/{2}$, both implying into the same results.

\begin{figure}[!htb]
			\subfloat[]{\label{figure41}\includegraphics[scale=0.63]{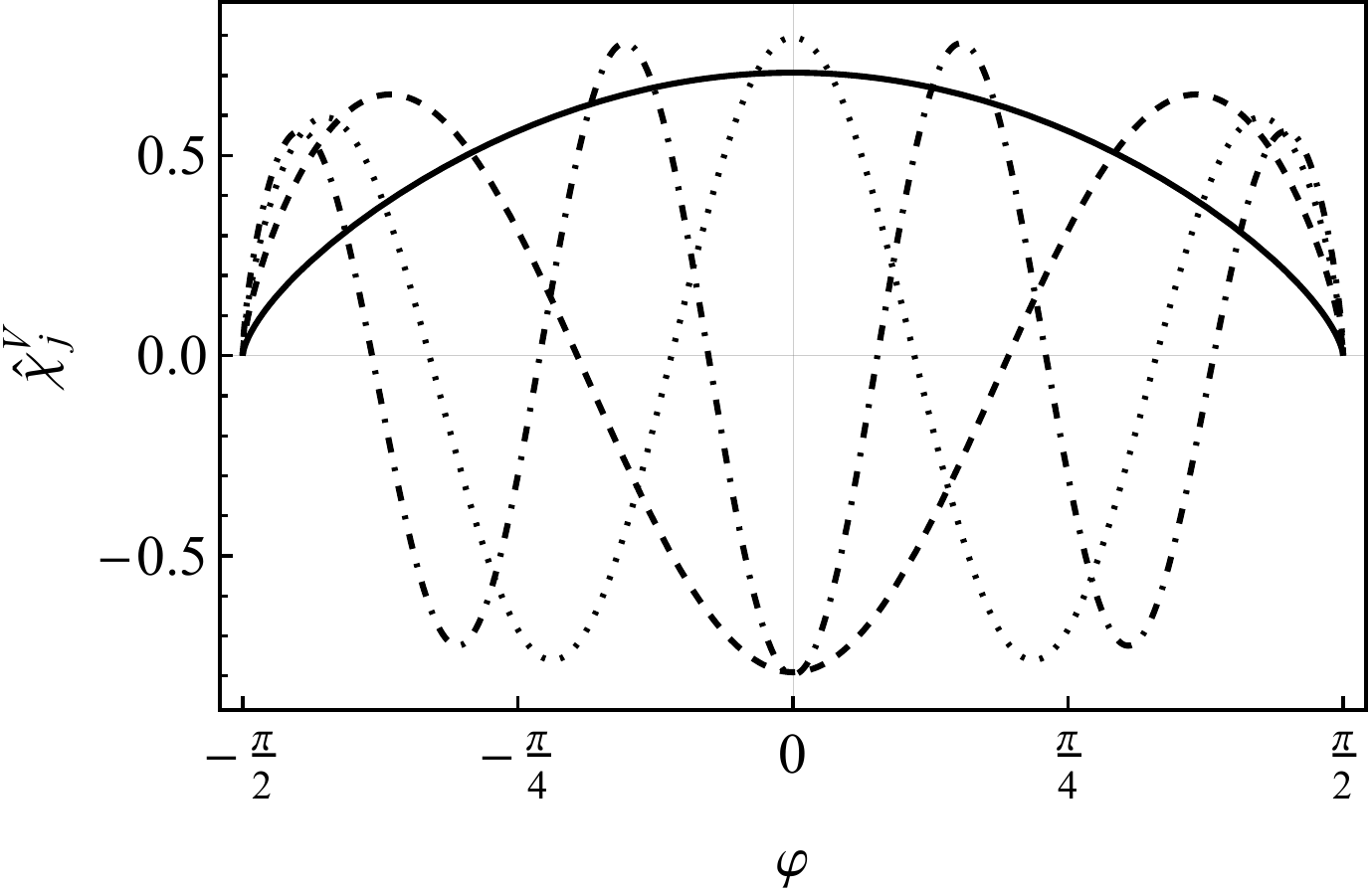}}$\;\;$
			\subfloat[]{\label{figure42}\includegraphics[scale=0.63]{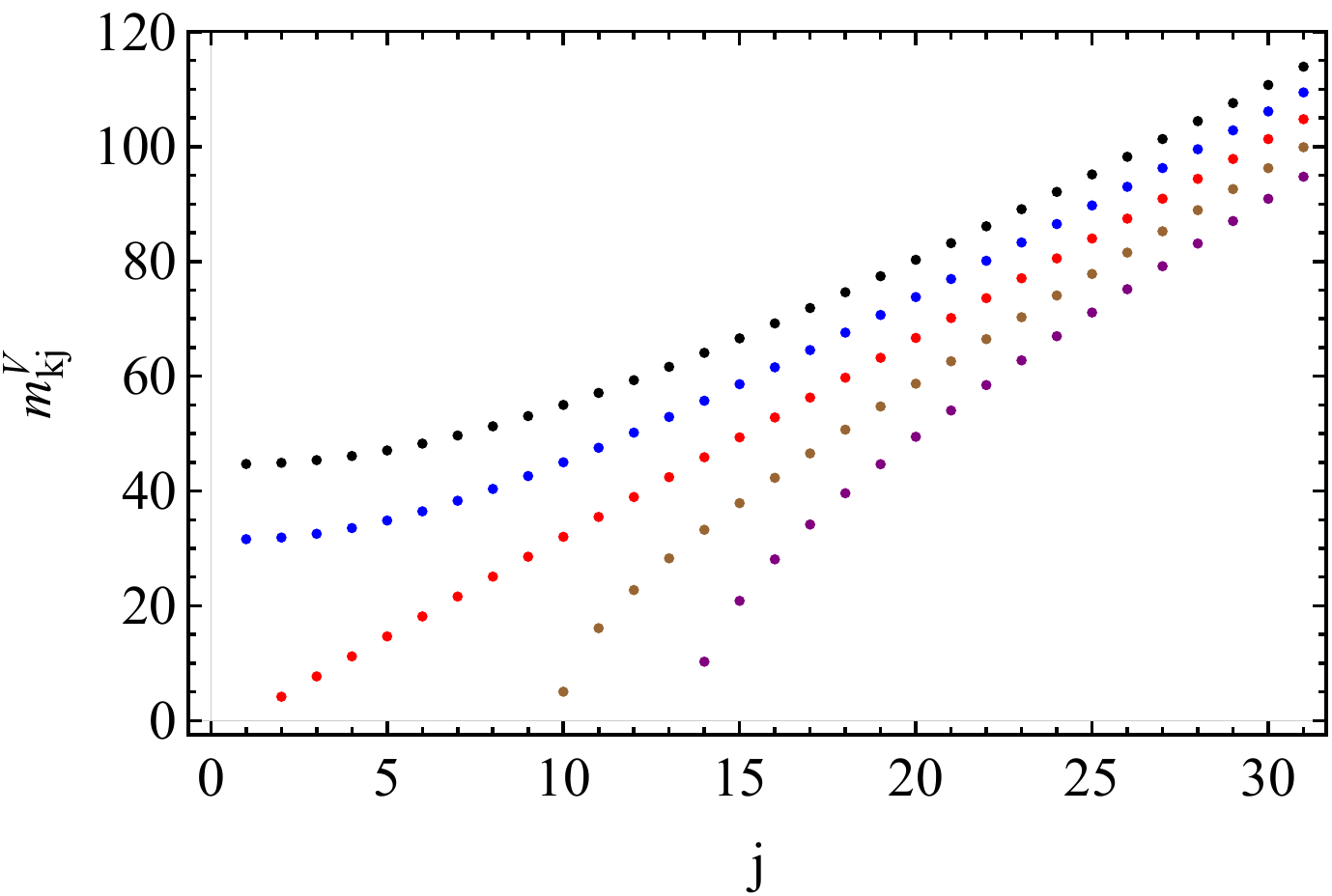}}
			
			\caption{(Color online) (a) Normalized wave functions $\hat{\chi}^{V}_{j}$ of model $V$ as functions of $\varphi=v\sqrt{3\left|\Lambda\right|}$ for $j=0$ (full line), $j=1$ (dashed line), $j=2$ (dotted line) and $j=3$ (dot-dashed line). (b) The graviton masses $m^{V}_{kj}$ of model V, with $\Lambda=-1$, for $k=-2\times10^{3}$ (black dots), $k=-10^{3}$ (blue dots), $k=0$ (red dots), $k=10^{3}$ (brown dots) and $k=2\times10^{3}$ (purple dots).}
		\end{figure}

The normalization condition for model $V$ reads
	\begin{equation}
	\int\left( e^{2\hat{A}-\hat{h}}\hat{\chi}_{mk}{}^{2} \right)\mathrm{d}v=\frac{\left(B^{V}_{j}\right)^{2}}{\sqrt{3\left|\Lambda\right|}}\int_{-\frac{\pi}{2}}^{\frac{\pi}{2}}\left[\sum_{i=0}^{j}\left(-1\right)^{i}\frac{j!}{\left(i!\right)^{2}\left(j-i\right)!}\frac{\Gamma\left(\frac{1}{2}+j+i\right)}{\Gamma\left(\frac{1}{2}+j\right)}\cos^{2i}\left(x\right)\right]^{2}\cos\left(x\right)\mathrm{d}x,
	\end{equation}
which is clearly finite, since the integrand is bounded at the entire domain $\left(-\pi/2,\pi/2\right)$ for every $j\in\mathbb{N}$. The normalized profile of the wave function, \eqref{stateV}, is depicted in Fig. \ref{figure41} and, for some particular values of $k$, the graviton masses are depicted in Fig. \ref{figure42}.

Notice that the constant $k$ is defined out of the eigenvalue spectrum of Eq.~\eqref{schroedingeru}, which is exactly the same Schr\"odinger-like equation, with energy $-k$, of the well known five dimensional braneworld models \cite{RS-1,RS-2,Bernardini2016,Almeida2009,Bazeia2004,Dzhunushaliev2010,Bazeia2002,DeWolfe2000,Ahmed2013,Gremm2000a,Gremm2000,Kehagias2001,Kobayashi2002,Bronnikov2003,BAZEIA2009b,BarbosaCendejas2013,ZHANG2008,Melfo2003,Bazeia2009a,Koley2005,Bazeia2014,Chinaglia2016,Bernardini2013}.

Before discussing sphere models, a short note about occurrence of instabilities in the above discussed models is pertinent. As identified, the scalar field perturbation is completely decoupled from tensorial ones, thus it be addressed separately. If the potential for the scalar fields do not present a local minima, occurrence of instabilities is not discarded.
However, the dynamics of the scalar field is not only governed by the potential, but also by the curvature of space, i.e. $A$, $h$ and $f$. Thus, even if there not being local minima from $\mathcal{V}$, stable configurations are possible under particular conditions. 
As an example, consider models $III$, $IV$ and $V$ \cite{Gauy2022}. The scalar field $\phi$ follows a similar structure to the ones found in five dimensional models \cite{Gauy2022,Gremm2000,Afonso2006,Sasakura2002}, thus should be stable by extension. Since model $IV$ only presents one scalar field ($\phi$) it should be stable. On the other hand, scalar field $\zeta$ could exhibit instabilities in models $III$ and $V$, since potential $\mathcal{V}$ is independent on $\zeta$. But the existence of curvature implies in the constraint equation,
\begin{equation}
\square\left(\zeta+\delta\zeta\right)=0\implies 
\hat{\square}\left(\delta\zeta\right)-4\,\hat{\sigma}^{ij}A_{,i}\left(\delta\zeta\right)_{,j}=0.\label{symmetricperturbation2}
\end{equation}
Therefore, the scalar field $\zeta$ satisfies the same stability equation (with $\Lambda=0$) as the gravitational field (cf. Eq.~\eqref{symmetricperturbation}), and should also be stable\footnote{Similarly, the stability analysis for models $I$ and $II$ is more intricate, since the potential depends on both $\phi$ and $\zeta$.}.

	\subsection{The $\mathbb{S}^{2}$ Models}\label{spheremodels}
	The sphere models based on models $III$ and $IV$ \cite{Gauy2022} are achieved by assuming that $\tilde{A}=-\ln\left(\sin\theta\right)$, $\tilde{f}=0$, $u=r\theta$ and $v=r\varphi$. Therefore, Eq.~\eqref{schroedingeru} can be written as
	\begin{equation}
	-\frac{\left[\sin\left(\theta\right) \tilde{\chi}^{\zeta}_{k,\theta}\right]_{,\theta}}{\sin\left(\theta\right)}+\left[\frac{4+kr^{2}}{\sin^{2}\left(\theta\right)}-6\right]\tilde{\chi}^{\zeta}_{k}=0,
	\end{equation}
	which has the solution
	\begin{equation}
	\tilde{\chi}^{\zeta}_{k}=B_{1k} P_2^{\sqrt{k r^2+4}}\left(\cos \left(\theta\right)\right)+B_{2k} Q_2^{\sqrt{k r^2+4}}\left(\cos \left(\theta\right)\right),\label{legendre}
	\end{equation}
	where $B_{1k}$ and $B_{2k}$ are normalization constants.

\begin{figure}[!htb]
		\includegraphics[scale=0.69]{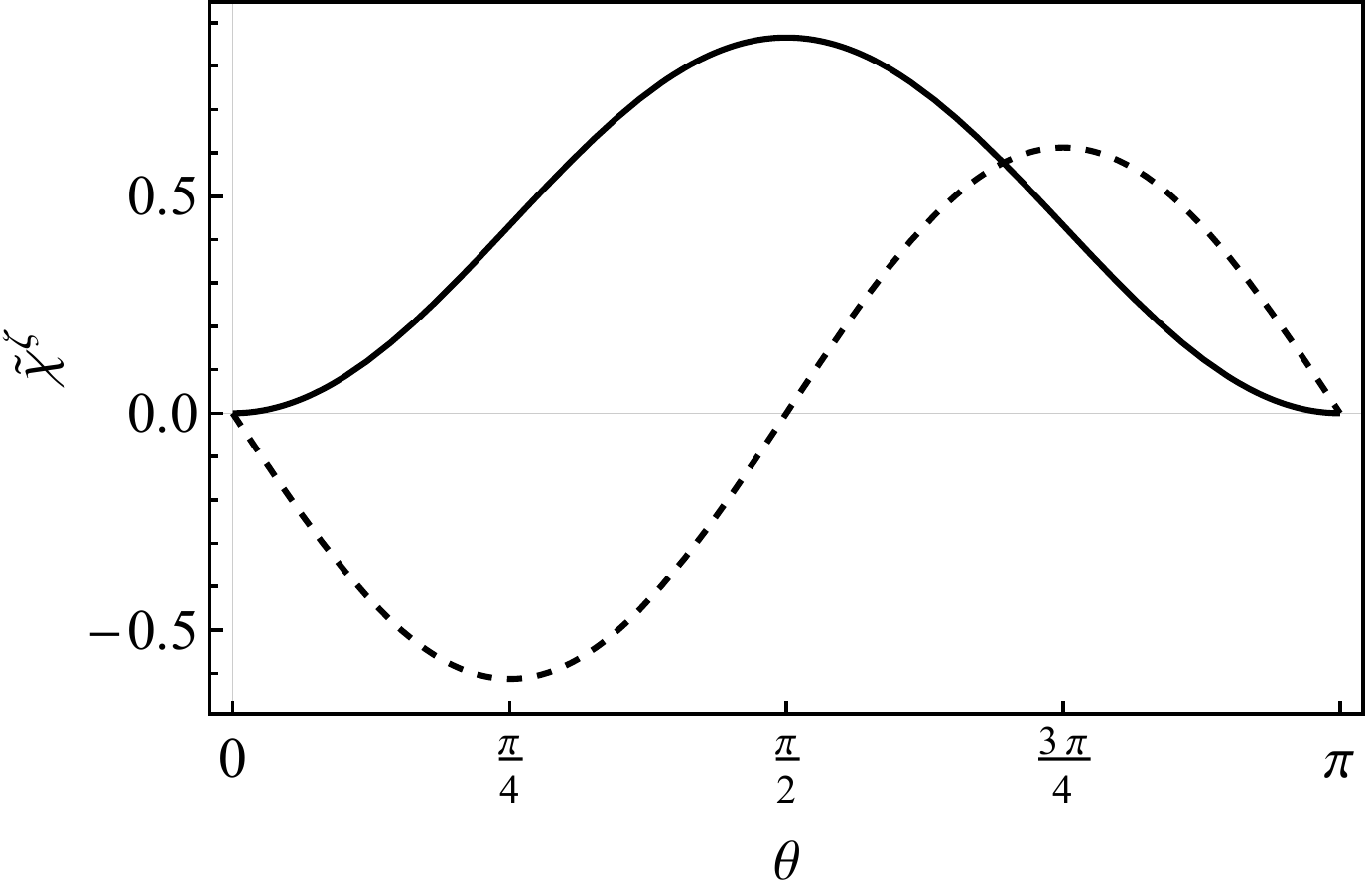}$\;\;$
		\caption{(Color online) Normalized gravitational wave functions $\tilde{\chi}^{\zeta}_{k}$ for the sphere models, for $k=0$ (full black line) and $k=-3/r^{2}$ (black dashed line).}\label{modes_sphere}
	\end{figure}

	To avoid singularities and trivial null solutions (which would be non-normalizable), expression \eqref{legendre} restricts the possible values of $k$. More precisely, normalizable modes  exclusively achieved for $k={-3}/{r^{2}}$ or $k=0$ are both depicted in Fig. \ref{modes_sphere}. However, the constant $B_{2k}$ must be fixed to zero, since $Q_2^{\sqrt{k r^2+4}}$ is non-normalizable in this space for all $k$.

\subsubsection{Model $III$ $(C=0)$}
Let the sphere models be constructed out of model $III$, with $C=0$, which is supported by the metric
\begin{equation}
\boldsymbol{g}^{III}_{\zeta}= \sin^{2}\left(\theta\right)\left(\eta_{\mu\nu}\mathrm{d}x^{\mu}\mathrm{d}x^{\nu}+r^{2}\mathrm{d}\varphi^{2}\right)+r^{2}\mathrm{d}\theta^{2}.
\end{equation}
The gravitational massive modes are then identified by
\begin{equation}
\chi^{III}_{\frac{\sqrt{n^{2}+3}}{r}}=B_{n1}\cos\left(n\varphi\right)\cos\left(\theta\right)\sin\left(\theta\right)\text{, } n\neq1;\label{trivalmode1}
\end{equation}
\begin{equation}
\chi^{III}_{\frac{n}{r}}=B_{n2}\cos\left(n\varphi\right) \sin^{2}\left(\theta\right)\text{, }n\neq2;
\end{equation}
\begin{equation}
\chi^{III}_{\frac{2}{r}}=B_{22}\cos\left(2\varphi\right)  \sin^{2}\left(\theta\right)+B_{11}\cos\left(\varphi\right) \cos\left(\theta\right)\sin\left(\theta\right).
\end{equation}

In this case, a point-like source, placed at $\left(\theta=\pi/2,\varphi=0\right)$, produces a Newtonian potential\footnote{Here one assumes that $B_{11}=0$ for simpler equations.} of the form
\begin{equation}
\phi^{III}_{N}(\mathsf{r})=\frac{3\mathcal{M}}{2^{8}\pi^{2}r^{2} M^{4}}\frac{1}{ \mathsf{r}}\left(1+\frac{2}{e^{\frac{\mathsf{r}}{r} }-1}\right)\approx\begin{cases}
\displaystyle\frac{3\mathcal{M}}{2^{8}\pi^{2}r^{2} M^{4}}\frac{1}{ \mathsf{r}}\text{, if }&\mathsf{r}\gg r
\\
\hfill
\\
\displaystyle\frac{3\mathcal{M}}{2^{7}\pi^{2}r M^{4}}\frac{1}{\mathsf{r}^{2}}\text{, if }&\mathsf{r}\ll r
\end{cases}.\label{newtonianIII}
\end{equation}

 Thus, in the region $\mathsf{r}\gg r$, the corrections to the Newtonian potential are exponentially suppressed, and one recovers the ordinary Newtonian theory.
Likewise, whenever $\mathsf{r}\ll r$, the result from \eqref{newtonianIII} is dominated by the higher dimensional term, thus gravity behaves as if the universe was five dimensional.
In summary, the Newtonian potential behaves exactly like in a five dimensional ADD model \cite{ArkaniHamed1998a}. On the other hand, the Newtonian gravitational constant is
\begin{equation}
G^{III}_{N}=\frac{3}{2^{8}\pi^{2}r^{2} M^{4}},\label{constantIII}
\end{equation}
which is similar to the Newtonian constant as determined from a six dimensional ADD model \cite{ArkaniHamed1998a}.

\subsubsection{Model $IV$}

Model $IV$ with a spherical internal space is characterized by the metric \cite{Gauy2022}:
\begin{equation}
\boldsymbol{g}^{IV_{n}}_{\zeta}= \frac{4r^{2}\Lambda}{3n^{2}}\cos^{2}\left(\frac{n \varphi}{2}\right) \sin^{2}\left(\theta\right)\omega^{+}_{\mu\nu}\mathrm{d}x^{\mu}\mathrm{d}x^{\nu}+r^{2}\mathrm{d}\theta^{2}+r^{2}\sin^{2}\left(\theta\right)\mathrm{d}\varphi^{2},
\end{equation}
where $n\in\left\{1,2\right\}$ and represents two distinct configurations of spherical models, which shall be labeled $IV_{1}$ and $IV_{2}$, for $n=1$ and $n=2$, respectively.

Correspondently, two normalizable graviton modes are identified,
\begin{equation}
\chi^{IV_{1}}_{\frac{2\Lambda}{3}}= B_{1}  \cos ^{2}\left(\frac{\varphi}{2}\right)\sin^{2}\left(\theta\right),
\end{equation}
and
\begin{equation}
\chi^{IV_{1}}_{\frac{4\Lambda}{3}}= B_{2}\sin\left(\frac{\varphi}{2}\right)  \cos ^{2}\left(\frac{\varphi}{2}\right)\sin^{2}\left(\theta\right),\label{wavesphere1}
\end{equation}
where $B_{1}$ and $B_{2}$ are normalization constants, and $\chi^{IV_{1}}_{\frac{2\Lambda}{3}}$ and $\chi^{IV_{1}}_{\frac{4\Lambda}{3}}$ represent gravitons of mass $m=\sqrt{2\Lambda/3}$ and $m=2\sqrt{\Lambda/3}$, respectively. The Newtonian potential associated with model $IV_{1}$ is
\begin{equation}
\phi^{IV_{1}}_{N}(\mathsf{r})=\frac{9\mathcal{M}}{2^{9}r^{4}\Lambda\,\pi^{2} M^{4}}\frac{e^{-\sqrt{\frac{2\Lambda}{3}}\mathsf{r}}}{ \mathsf{r}}\approx\frac{9\mathcal{M}}{2^{9}r^{4}\Lambda\,\pi^{2} M^{4}}\frac{1}{ \mathsf{r}},
\end{equation}
where the last approximation reinforces our choice of displacements much smaller than the cosmological scale, i.e. $\mathsf{r}\ll 1/\sqrt{\Lambda}$. Therefore, as far as the approximations here considered, model $IV_{1}$ implies, precisely, in Newton's theory, with the gravitational constant
\begin{equation}
G^{IV_{1}}_{N}=\frac{9}{2^{9}r^{4}\Lambda\,\pi^{2} M^{4}},
\end{equation}
which is dependent upon the cosmological constant $\Lambda$ and the radius $r$ of $\mathbb{S}^{2}$.

On the other hand, a spherical model with $n=2$ ($IV_{2}$) implies into two normalizable gravitational modes:
\begin{equation}
{\chi}^{IV_{2}}_{\frac{2\Lambda}{3}}=  B_{1}\cos ^{2}\left(\varphi\right) \sin^{2}\left(\theta\right),
\end{equation}
and
\begin{equation}
{\chi}^{IV_{2}}_{\frac{4\Lambda}{3}}= B_{2}\sin\left(\varphi\right)  \cos ^{2}\left(\varphi\right)\sin^{2}\left(\theta\right)+B_{3} \cos\left(\varphi\right)\sin\left(\theta\right)\cos\left(\theta\right) ,\label{wavesphere2}
\end{equation}
where $B_{1}$, $B_{2}$ and $B_{3}$ are normalization constants. 

The Newtonian potential for such a configuration is expressed by
\begin{equation}
\phi^{IV_{2}}_{N}(\mathsf{r})=\frac{9\mathcal{M}}{2^{7}\pi^{2} M^{4}r^{4}\Lambda}\frac{e^{-\sqrt{\frac{2\Lambda}{3}}\mathsf{r}}}{ \mathsf{r}}\approx\frac{9\mathcal{M}}{2^{7}\pi^{2} M^{4}r^{4}\Lambda}\frac{1}{ \mathsf{r}},
\end{equation}
where, once again, displacement are not of cosmological scale. Therefore model $IV_{2}$ also implies in Newtonian theory, with the gravitational constant
\begin{equation}
	G^{IV_{2}}_{N}=\frac{9}{2^{7}\pi^{2} M^{4}r^{4}\Lambda},
\end{equation}
which is a similar result to that one from model $IV_{1}$.

Had one proposed a more realistic model for matter\footnote{Which is outside the scope of the present paper.}, then Eqs. \eqref{wavesphere1} and \eqref{wavesphere2} would have significant effects in both the gravitational constant and Newtonian potential, which could imply into significant phenomenological differences between models $IV_{1}$ and $IV_{2}$. The same is true for model $III$: in a more realistic description, Eq. \eqref{trivalmode1} would modify Eqs. \eqref{newtonianIII} and \eqref{constantIII}.

\section{Conclusion}\label{Concl}

The gravitational fluctuations and phenomenological implications of some novel braneworld models (cf. Ref.~\cite{Gauy2022}) have been explored. For the so-called models $I$, $II$ and $III$, zero modes have been computed, and for models $IV$ and $V$, the entire spectrum of graviton modes were analytically identified. In addition, the implications of spherical internal spaces have been investigated for models $III$ and $IV$.

Generically, in contrast to ordinary five dimensional models, six dimensional models rely on a more intricate quantum mechanical two dimensional analogue problem, where the assessment of the localization of gravity is driven by the curved space co-dimensional variables, $(\mathbb{B}^{2},\boldsymbol{\hat{\sigma}})$, rendering a Schr\"odinger-like equation for each co-dimension, from which localization modes have been identified. 

Despite the reiterated physical appealing limitations of models $I$ and $II$ (cf. Ref.~\cite{Gauy2022}), the profile of the gravitational constant $G_{N}$ has been obtained, at least, as a pedagogical exercise. Results from model $I$ attest that the gravitational strength may be as small as one wishes by setting the parameter $p$ close to $3$, instead of reflecting some usual brane ``radius'' (parameters $c_{u}$ and $c_{v}$) tuning effect. In opposition, model $II$ behaves precisely as one would expect out of braneworlds: the gravitational strength is driven by the parameters $c_{u}$ and $c_{v}$.

More problematically, model $III$ implied into a Schr\"odinger-like equation whose massive states are not straightforwardly determined, with the exception for a trivial extension of five dimensional models ($C=0$), for which massive states are similar to flat compactified models. 

The most physically appealing intersecting braneworld, model $IV$, revealed a Schr\"odinger-like equation with a P\"osch-Teller potential, admitting a finite and discrete number of gravitons, whose masses are bounded from above by an interplay between the separation constant $k$ and the cosmological constant\footnote{For instance, if $k$ is non-positive one can find at most two massive gravitons.} $\Lambda$. Any propagating mode was summarily discarded after imposing unitary boundary conditions, thus rendering the singularities at the boundaries of space harmless. It is also noteworthy that model $IV$ presents a mass gap between the zero and massive modes, all the while retaining a smooth Ricci scalar but a singular Kretschmann scalar. This is in contrast to five dimensional models, where mass gaps and a singular Ricci scalar are directly connected \cite{HERRERAAGUILAR2010}. Thus we conjectured that the existence of a mass gap in higher dimensional braneworlds, other than five, may be connected to naked singularities of other scalar invariants besides the Ricci scalar.

The unique model constructed out of an anti-de Sitter brane, model $V$, has alluded to a Schr\"odinger-like equation with a trigonometric P\"osch-Teller potential and a discrete set of gravitational modes. Akin to the usual result in braneworlds, model $V$ presents countably, but infinite, gravitons whose masses are bounded from below by the value of the separation constant $k$.

Finally, concerning some technicalities pointed out by Ref.~\cite{Gauy2022}, through models $III$ and $IV$, one could also construct braneworld models whose internal space are spheres. The spherical models constructed out of model $III$, but with $C=0$, rely on a trivial extension of five dimensional braneworlds. Remarkably, such trivial extensions imply into a gravitational interaction that behaves like a five dimensional ADD model, while the gravitational constant behaves like a six dimensional one. On the other hand, sphere models constructed out of model $IV$ implied in two normalizable modes and the usual Newtonian interaction, with the gravitational constant being determined by a combination of the cosmological constant ($\Lambda$) and the radius of $\mathbb{S}^{2}$ ($r$).

Just to conclude, considering that the grounds for the above discussed intersecting braneworld models have been established, the localization of scalar, gauge and spinor fields shall be in the core of our subsequent investigations.

	\acknowledgments
	HMG is grateful for the financial support provided by
	CNPq (Grant No. 141924/2019-5). The work of AEB is supported by the Brazilian Agencies FAPESP (Grant No. 20/01976-5) and CNPq (Grant No. 301000/2019-0).

	\appendix

\section{On the decoupling between tensorial and scalar perturbations}\label{decoupling}
In Sec.~\ref{fluctuations}, the perturbations were cast form of Eq.~\eqref{metricperturbation}. Just as a preliminary  consideration, the general form from  Eq.~\eqref{metricperturbation} should involve perturbations of the warp factor, yet this can be disregarded because the latter can always be encompassed by $\varpi_{\mu\nu}$, as follows,
\begin{align}
\bar{\boldsymbol{g}}=& e^{-2\bar{A}}\left(\omega_{\mu\nu}+\varpi_{\mu\nu}\right)\mathsf{d}x^{\mu}\mathsf{d}x^{\nu}+\sigma_{ij}\mathsf{d}y^{i}\mathsf{d}y^{j}\label{metricapp}
\\
=& e^{-2A}\left(\omega_{\mu\nu}+\varpi_{\mu\nu}-2\delta A\,\omega_{\mu\nu}\right)\mathsf{d}x^{\mu}\mathsf{d}x^{\nu}+\sigma_{ij}\mathsf{d}y^{i}\mathsf{d}y^{j},
\end{align}
where $\bar{A}=A+\delta A$ and $\delta A$ is the perturbation of the warp factor. In which concerns the scalar perturbation related to $\delta A$, every calculation of Sec.~\ref{fluctuations} follows straightforwardly from defining $\tilde{\varpi}_{\mu\nu}=\varpi_{\mu\nu}-2\delta A\,\omega_{\mu\nu}$.

Of course, to clear up the above statement, from a more explicit approach, once that one assumes the metric to be \eqref{metricapp}, the gravitational portion of the action, if expanded up to the second order in the perturbative parameters, can be written as
\begin{align}
\bar{S}_{g}=& 2{M}^{4}\int\mathrm{d}^{6}x \sqrt{-\bar{\mathsf{g}}}\; \bar{R}\nonumber
\\
=& 2{M}^{4}\int\mathrm{d}^{6}xe^{-2\bar{A}}\sqrt{-\omega\sigma}\left[\mathcal{R}+e^{-2\bar{A}}\left(\Sigma-2\triangle^{2} \bar{A}-20\bar{A}^{,i}\bar{A}_{,i}\right)+10\hat{\square}\bar{A}-20\bar{A}^{,\mu}\bar{A}_{,\mu}-\varpi_{\mu\nu}\mathcal{R}^{\mu\nu}-20\varpi^{\mu\nu}A_{,\mu}\bar{A}_{,\nu}\right]\nonumber
\\
&+2{M}^{4}\int\mathrm{d}^{6}xe^{-2A}\sqrt{-\omega\sigma}\Bigg[\hat{\nabla}_{S}\left(\hat{g}^{SN}\varpi^{\mu\nu}\hat{\nabla}_{N}\varpi_{\mu\nu}-\hat{g}^{S\eta}\varpi^{\mu \nu}\hat{\nabla}_{\nu}\varpi_{\mu \eta}\right)-20{\varpi^{\mu}}_{\kappa}\varpi^{\kappa\nu}A_{,\mu}A_{,\nu}-\frac{1}{4}\varpi^{\mu\nu}\varpi_{\mu\nu}\hat{R}\nonumber
\Bigg.\\
&\qquad\qquad\qquad\qquad\qquad+10{\varpi^{\mu}}_{\kappa}\varpi^{\kappa \nu}\hat{\nabla}_{\mu}\hat{\nabla}_{\nu}A-\frac{1}{4}\hat{\nabla}^{K}\varpi^{\mu\nu}\hat{\nabla}_{K}\varpi_{\mu\nu}-\frac{5}{2}\varpi^{\mu\nu}\varpi_{\mu\nu}\hat{\square}A+5\varpi^{\mu\nu}\varpi_{\mu\nu}A^{,P}A_{,P}\nonumber
\\
&\qquad\qquad\qquad\qquad\qquad\qquad\quad\,\,\Bigg.+\frac{1}{2}\hat{\nabla}^{\nu}\varpi^{\mu \kappa}\hat{\nabla}_{\kappa}\varpi_{\mu \nu}+{\varpi^{\mu}}_{\kappa}\varpi^{\kappa\nu}\mathcal{R}_{\mu \nu}+5\hat{g}^{SJ}\varpi^{\mu \nu}\left(2\hat{\nabla}_{\nu}\varpi_{\mu J}-\hat{\nabla}_{J}\varpi_{\mu \nu}\right)A_{,S}\Bigg].
\end{align}
The only terms that can exhibit some coupling between the tensorial, $\varpi_{\mu\nu}$, and warp factor, $\delta A$, perturbations are\footnote{There are also $\varpi^{MN}\hat{\nabla}_{M}\hat{\nabla}_{N}\delta A$ and $\varpi_{MN}\hat{g}^{MN}\hat{g}^{AB}\hat{\nabla}_{A}\hat{\nabla}_{B}\delta A$, which are either null  $\left(\varpi_{MN}\hat{g}^{MN}\hat{g}^{AB}\hat{\nabla}_{A}\hat{\nabla}_{B}\delta A=0\right)$ or contribute only as a boundary term $\left(\varpi^{MN}\hat{\nabla}_{M}\hat{\nabla}_{N}\delta A=\hat{\nabla}_{M}\left(\varpi^{MN}\hat{\nabla}_{N}\delta A\right)\right)$ in the transverse traceless gauge ($\hat{\nabla}^{M}\varpi_{MN}=\hat{g}^{MN}\varpi_{MN}=0$). Therefore, one has a vanishing contribution from $\left(\partial_\mu \partial_\nu - \frac{1}{4} \omega_{\mu\nu} \omega^{\alpha\beta}\partial_\alpha \partial_\beta\right) \delta A$, i.e. the fluctuations of the scalar contributions are completely decoupled from the transverse traceless gravitational fluctuations, which is consistent with preliminary results in six dimensional models \cite{Csaki2000}.} $e^{-2\bar{A}}\varpi_{\mu\nu}\mathcal{R}^{\mu\nu}$ and $e^{-2\bar{A}}\varpi^{\mu\nu}{A}_{,\mu}\bar{A}_{,\nu}$. By construction, these terms are null, since $A_{,\mu}=\phi_{,\mu}=\zeta_{,\mu}=0$. Therefore, there is no coupling between $\varpi_{\mu\nu}$ and $\delta A$, such that neither $\varpi_{\mu\nu}$ nor $\delta A$ do affect one each other.

Likewise, the effects of the scalar fields ($\phi$ and $\zeta$) on the localization of gravitational fields can also be evaluated. For the metric described by Eq.~\eqref{metricperturbation}, the action of the scalar fields is written as
\begin{align}
\bar{S}_{\bar{\phi},\bar{\zeta}}=&-\int\mathrm{d}x^{6}\sqrt{-\bar{g}}\left[\frac{\bar{g}^{MN}}{2}\bar{\phi}_{,M}\bar{\phi}_{,N}+\frac{\bar{g}^{MN}}{2}\bar{\zeta}_{,M}\bar{\zeta}_{,N}+\mathcal{V}\left(\bar{\phi},\bar{\zeta}\right)\right]\nonumber
\\
=&-\int\mathrm{d}x^{6}\sqrt{-g}\left(1-\frac{1}{4}\varpi^{MN}\varpi_{MN}\right)\left[\frac{\bar{g}^{MN}}{2}\bar{\phi}_{,M}\bar{\phi}_{,N}+\frac{\bar{g}^{MN}}{2}\bar{\zeta}_{,M}\bar{\zeta}_{,N}+\mathcal{V}\left(\bar{\phi},\bar{\zeta}\right)\right]\nonumber
\\
=&\frac{1}{4}\int\mathrm{d}x^{6}\sqrt{-g}\,\varpi^{MN}\varpi_{MN}\left[\frac{g^{MN}}{2}\phi_{,M}\phi_{,N}+\frac{g^{MN}}{2}\zeta_{,M}\zeta_{,N}+\mathcal{V}\left(\phi,\zeta\right)\right]\nonumber
\\
&\qquad\qquad\qquad-\int\mathrm{d}x^{6}\sqrt{-g}\left[\frac{{g}^{MN}}{2}\bar{\phi}_{,M}\bar{\phi}_{,N}+\frac{{g}^{MN}}{2}\bar{\zeta}_{,M}\bar{\zeta}_{,N}+\frac{\varpi^{MN}}{2}\bar{\phi}_{,M}\bar{\phi}_{,N}+\frac{\varpi^{MN}}{2}\bar{\zeta}_{,M}\bar{\zeta}_{,N}+\mathcal{V}\left(\bar{\phi},\bar{\zeta}\right)\right],
\end{align}
where $\bar{\phi}=\phi+\delta\phi$ and $\bar{\zeta}=\zeta+\delta\zeta$. The only relevant terms for the coupling between tensorial and scalar perturbations are $\varpi^{MN}\bar{\phi}_{,M}\bar{\phi}_{,N}$ and $\varpi^{MN}\bar{\zeta}_{,M}\bar{\zeta}_{,N}$, which once again are evidently null since $\varpi^{MN}\phi_{,M}=\varpi^{MN}\zeta_{,M}=0$.

Hence, to summarize, the tensorial fluctuations, $\varpi_{\mu\nu}$, are completely decoupled from the scalar perturbations, i.e. $\delta A$, $\delta \phi$ and $\delta\zeta$, and the action that drives the dynamics of $\varpi_{\mu\nu}$ can be cast in the form of \eqref{actiontotal}, as pointed out in Ref.~\cite{Csaki2000}. In the above context, one can disregard the perturbations of the scalar fields when dealing with the dynamics of gravity.

\section{The point-like particle stress energy tensor}\label{ParticleStress}

From a general standpoint, one may write the action of matter fields as
\begin{equation}
S_{\mathcal{M}}=\int \mathrm{d}^{6}x \sqrt{-\bar{g}}\mathcal{L}_{\mathcal{M}}(\Phi,\bar{\boldsymbol{g}}),
\end{equation}
where $\Phi$ represents all matter fields. If $\bar{\boldsymbol{g}}=\boldsymbol{g}+\boldsymbol{\varpi}$, where $\boldsymbol{\varpi}$ is a perturbation, then the matter action to first order in $\varpi$ takes the form
\begin{equation}
S_{\mathcal{M}}=\int \mathrm{d}^{6}x \sqrt{-g}\mathcal{L}_{\mathcal{M}}\left(\Phi,\boldsymbol{g}\right)+\sum_{m\in\mathbb{I}}\int \mathrm{d}^{6}x e^{-6A}\sqrt{-\hat{g}}\;\;\chi_{m}\left({\widetilde{\varpi}_{m}}\right)_{\mu\nu}\frac{{T}^{\mu\nu}}{2},\label{firstorderaction}
\end{equation}
where ${T}^{\mu\nu}$ is the stress energy tensor calculated out of $\mathcal{L}_{\mathcal{M}}(\Phi,\boldsymbol{g})$. On the other hand, if the matter is regarded as a point-like particle of mass $\mathcal{M}$, then the stress energy tensor is,
\begin{equation}
S_p=\mathcal{M}\int\mathrm{d}\tau\sqrt{-{g}_{MN}\mathsf{v}^{M}\mathsf{v}^{N}}\implies T^{MN}(x^{Q})=\frac{\mathcal{M}\delta(x^{Q}-x^{P})\mathsf{v}^{M}(x^{P})\mathsf{v}^{N}(x^{P})}{\sqrt{-g}}.\label{particlestresstensor}
\end{equation}

Substituting \eqref{particlestresstensor} into \eqref{firstorderaction} reduces to \eqref{particlefirst}.

\section{On the smoothness of hypergeometric functions}\label{Discontinuities}
	
Assume a general wave function of the form
\begin{equation}
\chi=\cos^{l}\left(y\right)\,{}_{2}F_{1}\left[\alpha,\beta;\gamma;\cos^{2}\left(y\right)\right],
\end{equation}
its first derivative is simply
\begin{equation}
\chi_{,y}=-l\cos^{l-1}\left(y\right)\sin\left(y\right)\,{}_{2}F_{1}\left[\alpha,\beta;\gamma;\cos^{2}\left(y\right)\right]-2\sin\left(y\right)\cos^{l+1}\left(y\right)\,{}_{2}F_{1}'\left[\alpha,\beta;\gamma;\cos^{2}\left(y\right)\right].
\end{equation}

The derivatives of hypergeometric functions follow the rule
\begin{equation}
_{2}F_{1}'\left[\alpha,\beta;\gamma;z\right]=\frac{\mathrm{d}\;{}_{2}F_{1}\left[\alpha,\beta;\gamma;z\right]}{\mathrm{d}z}=\frac{\alpha\beta}{\gamma}\,{}_{2}F_{1}\left[\alpha+1,\beta+1;\gamma+1;z\right].
\end{equation}

Substituting into $\chi_{,y}$ one finds
\begin{equation}
\chi_{,y}=\cos^{l-1}\left(y\right)\sin\left(y\right)\left\{-l\,{}_{2}F_{1}\left[\alpha,\beta;\gamma;\cos^{2}\left(y\right)\right]-2\cos^{2}\left(y\right)\frac{\alpha\beta}{\gamma}\,{}_{2}F_{1}\left[\alpha+1,\beta+1;\gamma+1;\cos^{2}\left(y\right)\right]\right\}.
\end{equation}

Let the parameters $\alpha$, $\beta$ and $\gamma$ satisfy $\gamma-\alpha-\beta=1/2$, then ${}_{2}F_{1}\left[\alpha+1,\beta+1;\gamma+1;\cos^{2}\left(y\right)\right]$ will not converge at $\cos^{2}\left(y\right)=1$. To analyze this divergence further one uses the following property of hypergeometric functions \cite{Fluegge1998}:
\begin{align}
{}_{2}F_{1}\left(\alpha,\beta;\gamma;z\right)&=\frac{\Gamma(\gamma)\Gamma(\gamma-\alpha-\beta)}{\Gamma(\gamma-\alpha)\Gamma(\gamma-\beta)}{}_{2}F_{1}\left(\alpha,\beta;\alpha+\beta-\gamma+1;1-z\right)\nonumber
\\
&\qquad\qquad+\frac{\Gamma(\gamma)\Gamma(\alpha+\beta-\gamma)}{\Gamma(\alpha)\Gamma(\beta)}\left(1-z\right)^{\gamma-\alpha-\beta}{}_{2}F_{1}\left(\gamma-\alpha,\gamma-\beta;\gamma-\alpha-\beta+1;1-z\right),
\end{align}
thus the first derivative satisfy
\begin{align}
\chi_{,y}=&-l\cos^{l-1}\left(y\right)\sin\left(y\right)\,{}_{2}F_{1}\left[\alpha,\beta;\gamma;\cos^{2}\left(y\right)\right]\nonumber
\\
&\qquad\qquad-2\cos^{l+1}\left(y\right)\sin\left(y\right)\frac{\alpha\beta}{\gamma}\frac{\Gamma(\gamma+1)\Gamma(-1/2)}{\Gamma(\gamma-\alpha)\Gamma(\gamma-\beta)}{}_{2}F_{1}\left[\alpha+1;\beta+1,3/2;\sin^{2}\left(y\right)\right]\nonumber
\\
&\qquad\qquad\qquad\qquad-2\cos^{l+1}\left(y\right)\sin\left(y\right)\frac{\alpha\beta}{\gamma}\frac{\Gamma(\gamma+1)\Gamma(1/2)}{\Gamma(\alpha+1)\Gamma(\beta+1)}\left[\sin^{2}\left(y\right)\right]^{-1/2}{}_{2}F_{1}\left[\gamma-\alpha,\gamma-\beta;1/2;\sin^{2}\left(y\right)\right],
\end{align}
and at the vicinity of $y=0$ it behaves like
\begin{equation}
\chi_{,y}\Big|_{y\rightarrow0}=-2\,\text{sign}\left[\sin\left(y\right)\right]\frac{\alpha\beta}{\gamma}\frac{\Gamma(\gamma+1)\Gamma(1/2)}{\Gamma(\alpha+1)\Gamma(\beta+1)},
\end{equation}
leading to a discontinuity unless $\alpha$ or $\beta$ is a non-positive integer $-j$, $j\in\mathbb{N}$.

	\footnotesize
	

\end{document}